\documentclass{aa}  
\usepackage[svgnames]{xcolor}  
\usepackage{graphicx}
\usepackage{natbib}
\usepackage{natbib,twoopt}
\usepackage[breaklinks=true,colorlinks,citecolor=blue]{hyperref} 
\bibpunct{(}{)}{;}{a}{}{,} 
\makeatletter
\newcommandtwoopt{\citeads}[3][][]{\href{http://adsabs.harvard.edu/abs/#3}%
  {\def\hyper@linkstart##1##2{}%
  \let\hyper@linkend\@empty\citealp[#1][#2]{#3}}}
\newcommandtwoopt{\citepads}[3][][]{\href{http://adsabs.harvard.edu/abs/#3}%
  {\def\hyper@linkstart##1##2{}%
  \let\hyper@linkend\@empty\citep[#1][#2]{#3}}}
\newcommandtwoopt{\citetads}[3][][]{\href{http://adsabs.harvard.edu/abs/#3}%
  {\def\hyper@linkstart##1##2{}%
  \let\hyper@linkend\@empty\citet[#1][#2]{#3}}}
\newcommandtwoopt{\citeyearads}[3][][]%
  {\href{http://adsabs.harvard.edu/abs/#3}
  {\def\hyper@linkstart##1##2{}%
  \let\hyper@linkend\@empty\citeyear[#1][#2]{#3}}}
\makeatother

\hypersetup{
    colorlinks = true,
    linkcolor = blue,
    anchorcolor = blue,
    citecolor = blue,
    filecolor = blue,
    urlcolor = blue
    }

\usepackage{xspace}
\usepackage{amsmath}

\newcommand{\spec}[3]{\mbox{#1\,{\sc #2}\,$\lambda$#3}\xspace}

\usepackage{txfonts}
\begin{document} 

   \title{Newborn Be star systems observed shortly after mass transfer}


   \author{Th.\ Rivinius
          \inst{1}
          \and
          R.\ Klement
          \inst{1,2}
          \and
          S.~D.\ Chojnowski
          \inst{3}
          \and
          D.\ Baade
          \inst{4}
          \and
          M.\ Abdul-Masih
          \inst{1}
          \and
          N.\ Przybilla
          \inst{5}
          \and
          J.\ Guarro Fl\'o
          \inst{6}
          \and
          B.~Heathcote
          \inst{7}
          \and
          P.\ Hadrava
          \inst{8}
          \and
          D.\ Gies
          \inst{9}
          \and
          K.\ Shepard
          \inst{9}
          \and
          C.\ Buil
          \inst{10}
          \and
          O.\ Garde
          \inst{11}
          \and
          O.\ Thizy
          \inst{12}
          \and
          J.\ D.\ Monnier
          \inst{13}
          \and
          N.\ Anugu
          \inst{2}
          \and
          C.~Lanthermann
          \inst{2}
          \and
          G.\ Schaefer
          \inst{2}
          \and 
          C.\ Davies
          \inst{14}
          \and
          S.\ Kraus
          \inst{14}
          \and
          J.\ Ennis
          \inst{13}
          \and
          B.\ R.\ Setterholm
          \inst{13}
          \and
          T.\ Gardner
          \inst{14}
          \and
          N.\ Ibrahim
          \inst{13}
          \and
          S.~Chhabra
          \inst{14}
          \and
          M.\ Gutierrez
          \inst{13}
          \and
          I.\ Codron
          \inst{14}
          }

   \institute{European Organisation for Astronomical Research in the Southern Hemisphere (ESO), Casilla 19001, Santiago 19, Chile\\
              \email{triviniu@eso.org}
         \and The CHARA Array of Georgia State University, Mount Wilson Observatory, Mount Wilson, CA 91023, USA
         \and NASA Ames Research Center, Moffett Field, CA 94035, USA
         \and European Organisation for Astronomical Research in the Southern Hemisphere (ESO), Karl-Schwarzschild-Str.\ 2, 85748 Garching b.\ München, Germany
         \and Universit\"at Innsbruck, Institut f\"ur Astro- und Teilchenphysik, Technikerstr. 25/8, 6020 Innsbruck, Austria
         \and Aras Observers Group, Carrer Balmes, 2.\ 08784 Piera (Barcelona), Spain
         \and 269 Domain Road, Melbourne, Victoria, 3141, Australia 
         \and Astronomical Institute, Academy of Sciences of the Czech Republic, Bo\v{c}n\'{\i} II 1401, CZ-14100 Prague, Czech Republic
         \and  Center for High Angular Resolution Astronomy and Department of Physics and Astronomy, Georgia State University, P.O. Box 5060, Atlanta, GA 30302-5060, USA
         \and Observatoire Antibes Saint-Jean, 91 Avenue Francisque Perraud, 06600 Antibes, France
         \and Observatoire de la Tourbi\'ere, 45 Chemin du Lac, 38690 Ch\^abons, France
         \and Observatoire Belle-Etoile, 250 route de la Belle Etoile, 38420 Revel, France
         \and Astronomy Department, University of Michigan, Ann Arbor, MI 48109, USA
         \and Astrophysics Group, Department of Physics \& Astronomy, University of Exeter, Stocker Road, Exeter, EX4 4QL, UK
             }

   \date{Received xxx; accepted yyy}

 
  \abstract
   {{Many classical Be stars acquire their very rapid rotation} by mass and angular-momentum transfer in massive binaries, marking the first phase of the evolutionary chain. {Later-stage products}, such as Be+subdwarf and Be+neutron-star binaries (Be X-ray binaries), are also well known, although the search and definite proof for Be+white dwarf companions is ongoing.  Short-lived intermediate-phase objects, i.e., binaries past the interaction stage but with the donor stars not yet having reached the end of their evolution/contraction, have only been discovered recently.   } 
   {The main hallmark of the first such binaries of this kind is a system of absorption lines with low width, significant radial-velocity variations, and peculiar relative line strengths.  Data archives and the literature can be searched for additional candidates exhibiting this pattern, and follow-up observations be obtained in order to increase the number of these systems with quantitatively known orbits to provide a basis for an initial statistical investigation and to develop observational strategies for abundance analyses.}
   {Thirteen candidates were identified at various confidence levels.  To verify the candidates' nature, orbital elements were derived from new high-quality spectra and interferometric observations where possible, qualitative analyses of other basic parameters were performed, and indicators of advanced stages of nucleosynthesis were preliminarily evaluated.}
   {Adding to the two known systems identified as classical Be star+pre-subdwarf binaries (LB-1 and \object{HR\,6819}), two more (\object{V742\,Cas}, \object{HD\,44637}) could be confirmed with interferometry, with V742\,Cas setting a new record for the smallest visually observed angular semi-major axis, at $a=0.663$\,mas. Two further ones (\object{V447\,Sct}, \object{V1362\,Cyg}) are not resolved interferometrically, {but other evidence} puts them at the same confidence level as LB-1.  \object{V2174\,Cyg} is a candidate with very high confidence, but was not observed interferometrically. The remaining ones are either candidates with varying levels of confidence, mainly so due to the lack of available spectroscopic or interferometric observations for comparison with the others and orbit determination, or could be rejected as candidates with the followup observations.
 }
   {Of a mostly magnitude complete sample of 328 Be stars, {0.5--1\%} are found to have recently completed the mass overflow that led to their formation. Another 5\% are systems with compact subdwarf companions, i.e., further evolved after a previous overflow, and possibly two more percent harbor white dwarfs. All these systems are of early B-subtypes, however, and if the original sample is restricted to early subtypes {(136 objects)}, these percentages increase by a factor of about 2.5, {while dropping to zero for the mid and late subtypes (together 204 objects)}. This strongly suggests that early- vs. mid- and late-type Be stars have differently weighted channels to acquire their rapid rotation, namely binary interaction vs.\ evolutionary spin-up.}
   

   \maketitle
%
%

\section{Introduction}\label{sec:intro}
The classical B emission line stars, {or Be stars}, are typically understood as rapidly rotating B-type stars, in which an additional mechanism acts to eject photospheric material, forming a gaseous Keplerian, slowly outflowing viscous decretion disk (VDD). Knowledge about the disk and its physics has seen significant advances over the past decades \citep[as summarized by][]{2013A&ARv..21...69R}. The "additional mechanism" is largely suspected to be related to pulsation (op cit., although that is not quite an unequivocal opinion, see \citealt{2021ApJ...921....5B}). The origin of the rapid rotation, though, has been suggested to lie either in the interior evolution of a single star, via angular momentum transport from the contracting core to the surface \citep[see, e.g.,][]{2013A&A...553A..25G}, or in binary evolution, where the rotation is the {product of mass-transfer} between a more rapidly evolving component and a mass gainer, that is observed as a Be star. The former case could be proven, for instance, by identifying main sequence companions to Be stars in orbits too close to accommodate a previous mass-transfer (\citealt{2020A&A...641A..42B} {find no such systems among early-type Be stars, but see also \citealt{2022A&A...667A.111K}, who suggest Achernar, although being of later subtype  than the cutoff of \citeauthor{2020A&A...641A..42B}, is one case}), or through statistical means. The latter case would typically either show properties such as anomalous space velocities, in case the former donor had exploded and the system been disrupted
\citepads{2001ApJ...555..364B}, or leave behind the remnant of the previous mass-donor. Such a remnant, unless producing strong X-rays {through accretion on a compact object}, would, however, be difficult to detect as it would contribute only a low percentage to the total flux
\citep[see][for descriptions of spectral modeling and reports on the most recent search efforts]{
2018A&A...615A..78G,2022Ap&SS.367..124J,2022ApJ...926..213K,2023AJ....165..203W}.

There is one short period of time, though, in which such a remnant could be easily seen next to the Be star, and possibly even outshine it. That is when the mass transfer has fully completed, and the Be star is already a classical Be star in the above sense (i.e., it is producing its own decretion disk through mass ejection), but the former donor has not yet contracted to its later equilibrium configuration of a hot subdwarf and could be dubbed a pre-subdwarf.  In this case, the spectral appearance might be dominated by the photospheric spectrum of the former donor, and show some quite unique features \citepads{2022A&A...667A.122S,2022NatAs...6.1414I}. Such subdwarfs would later go on to become white dwarfs, but in the context of this work only the immediate, {post-overflow} subdwarf phase is of concern. {Be+WD systems relevant to the binarity connected evolution and formation of Be stars could be formed via another channel \citep[case BB mass transfer, as described by][]{2023ApJ...942L...6G} and are discussed only in Sect.~\ref{sec:discussion}}.

The current discussion on Be star binary systems that might have just completed mass transfer was triggered observationally, when \citetads{2019Natur.575..618L} initially suggested LB-1, also known as \object{ALS\,8775}, as a very massive stellar black hole with a Be star companion. This was quickly refuted, and of the many alternative hypotheses that of a classical Be star with a stripped and bloated post-Roche-lobe-overflow companion that has not yet contracted to its equilibrium radius and temperature has become the generally accepted one \citepads{2020AA...639L...6S}.

In the wake of the discovery of LB-1, \citetads{2020A&A...637L...3R} posited \object{HR\,6819} as a Be star with a quiescent stellar-mass black-hole companion as one of the alternative scenarios applicable also to LB-1. Yet, little later it was suggested to be a pre-subdwarf companion as well by \citetads{2020AA...641A..43B} and \citetads{2021MNRAS.502.3436E}. Using observations with the Very Large Telescope Interferometer (VLTI), this interpretation was proven correct by \citetads{2022AA...659L...3F}, who could determine an initial astrometric orbit of two similarly bright objects from the two first observations of an ongoing study.

{The field is evolving rapidly, and more objects are proposed frequently, such as \citet{2023A&A...674L..12R} and \citet{2023MNRAS.525.5121V}. Both systems are in the Magellanic Clouds, however, while the current study investigates the local environment of bright stars, making it much easier for follow-up studies to look at the reported objects in detail. }

Here we identify and describe five other systems that bear the properties of a stripped donor and spun-up companion star. In the following, Sect.~\ref{sec:obs} introduces the candidate selection criteria and the observations, on which this work is based, both our own and archival ones, while Sect.~\ref{sec:systems} discusses the systems consisting of a Be star and a confirmed or candidate post-overflow companion. In addition to the two cases introduced above, we present several systems ranging from an observationally confirmed nature to a moderately confident candidate level. 
\citetads{2022MNRAS.516.3602E} suggest \object{HD\,15124} to be an immediate progenitor of such systems, but as shown in Sect.~\ref{sec:discussion}, many such systems may be known already, as the system parameters of HD\,15124 do not seem to be particularly exceptional.
Section~\ref{sec:discussion} further considers statistics, the relation to the general paths of binary evolution as well as in what phases of stellar evolution and binary interaction the individual systems might be, without, however, attempting an exhaustive quantitative analysis, which is deferred to later works. The Appendices show spectroscopic data (\ref{app:spectroscopic}), interferometric data (\ref{app:inter}), and list additional Be stars with composite spectra and thus potential binaries that were found to be of interest, but do not quite fit into the main body (\ref{app:compspec}).

\section{Observational data}\label{sec:obs}
\subsection{Candidate identification}
\citetads{2002ASPC..279..143C} gave a compilation of six B-type binaries with invisible, but supposedly massive companions due to their high mass function. Considering the properties of LB-1 and HR\,6819, these are of particular interest, and indeed five of the companions turn out to be good candidates for stripped and bloated stars. Additional potential systems were identified by visually browsing archival data, as detailed below.  Again following the leads provided by LB-1 and HR\,6819, but also by examining other systems presented here, candidates were identified by looking for Be stars with absorption lines drastically narrower than would be expected from the emission line shape \citep{2013A&ARv..21...69R}. Further criteria were that they either show radial velocity (RV) variations in these absorption lines, but not in the emission, or potentially abnormal strength pattern in the spectral lines of H, He, and CNO. Spectra of candidates can be seen in Appendix~\ref{app:spectroscopic}.

{Considering the inhomogeneity of the sources from which the candidates were drawn, there is no statistically meaningful parent sample. Such a sample can, nevertheless, be constructed a posteriori, as will be shown in Sect.~\ref{sec:discussion}. }

\subsection{Interferometry}

Several of the program stars were observed using optical/near-IR interferometry with the goal of detecting the companions (see below for a more detailed description of the facilities used). In this work we present new interferometric observations for eight of the program stars: VLTI data for LB-1, HR\,2309, HR\,3195, HD\,44637, and V1371\,Tau, and CHARA Array data for V742~Cas, V447~Sct, and V1362~Cyg. Close, low-contrast companions were successfully detected for four stars -- V742\,Cas, HR\,2309, HD\,44637, and V1371\,Tau -- and the astrometric orbit of V742\,Cas was subsequently mapped with additional measurements. 

The log of the interferometric data can be found in Table~\ref{tab:interf_log}. The calibrator stars were selected using the Searchcal software developed and maintained by JMMC \footnote{\url{https://www.jmmc.fr/english/tools/proposal-preparation/search-cal/}} \citepads{2016A&A...589A.112C}. The calibrator angular diameters were adopted from the JMMC catalog of stellar diameters \citepads{2014ASPC..485..223B, 2017yCat.2346....0B} and are listed in Table~\ref{tab:calibrators}.

For interferometric data with low spectral resolution (LB-1, V742~Cas, V447~Sct, and V1362~Cyg), the \textit{CANDID} code was used to obtain (1) the relative positions and flux ratios of the companion in the case of successful detections, and (2) minimum magnitude differences for the unseen companions in case of non-detections. For details about the \textit{CANDID} code, the reader is referred to the dedicated work presenting the code \citepads{2015A&A...579A..68G} as well as more recent works that made use of it \citep[e.g.][]{2022ApJ...926..213K}. To facilitate easier manipulation of interferometric data with high spectral resolution (HR\,2309, HR\,3195, HD\,44637, and V1371\,Tau), the more versatile code for interferometric modeling \textit{PMOIRED} \citepads{2022SPIE12183E..1NM} was employed to achieve the same goals. Furthermore, high spectral resolution $K$-band data enable resolving the Br$\gamma$ line, which is useful to study kinematics of the circumstellar environment and to identify whether the emission-line star is the brighter or fainter component in a low-contrast binary. Figures in Appendix~\ref{app:inter} show examples of the data analysis.

\subsubsection{VLTI}

The Very Large Telescope Interferometer (VLTI) is a facility located on Cerro Paranal in Chile, that can be fed by either the four 8.2-m unit telescopes (UTs), or an array consisting of four movable 1.8-m auxiliary telescopes (ATs). The angular resolution, which is defined as the angular size of an object where visibility goes through the first null, will depend on the projected length of the baseline configuration used. For the VLTI with UTs, this is at most 4\,mas, and for the VLTI with the ATs put in large configuration it is at most 2\,mas. The instrument used here, GRAVITY, takes four telescope beams and combines them in the K-band, with a choice of spectral resolutions of $R\sim22\,500$ and 4000 \citepads{2017A&A...602A..94G}. Observations were reduced with the GRAVITY pipeline workflow provided under ESO {\it Reflex} \citepads{2013A&A...559A..96F}. The VLTI was used with the UTs for LB-1 in low spectral resolution, and with the ATs located in the astrometric baseline configurations for HR\,2309, HD\,44367, and HR\,3195, while for V1371\,Tau they were in the medium baseline configuration, in high spectral resolution. 

\subsubsection{CHARA Array}

The Center for High Angular Resolution Astronomy (CHARA) Array \citepads{2005ApJ...628..453T, 2020SPIE11446E..05S} is an optical/near-IR interferometer located on Mt.\ Wilson, California, USA, and consists of six 1-m telescopes in a Y-shaped configuration. The maximum baseline of $B_\mathrm{max} = 330$\,m yields an angular resolution of $\lambda / (2B_\mathrm{max}) \sim$0.5\,milliarcsec (mas) in the near-IR $H$ band and $\sim$0.7\,mas in the $K$ band. The six-telescope beam combiners MIRC-X \citepads{2020AJ....160..158A} and MYSTIC \citepads{Setterholm+2023}, operating simultaneously in the near-IR $H$ and $K$ bands, respectively, are capable of detecting binary companions with snapshot observations (on-source integration time of 10--20 minutes) down to a contrast of $\sim$0.5\%. The default spectral resolution of $R\sim$50 corresponds to an interferometric field of view (FoV) of $\sim$50 and $\sim$65\,mas for the two instruments, respectively. The data for V742~Cas, V447~Sct, and V1362~Cyg taken with the default settings were reduced and calibrated using dedicated software\footnote{\url{https://gitlab.chara.gsu.edu/lebouquj/mircx_pipeline.git}} \citep[pipeline version 1.3.5,][]{2020AJ....160..158A} and will be made publicly available in the Optical Interferometry Database\footnote{\url{http://oidb.jmmc.fr/index.html}} \citepads{2014SPIE.9146E..0OH} and the CHARA Data Archive\footnote{\url{https://www.chara.gsu.edu/observers/database}}.

\subsection{Spectroscopy}
\subsubsection{UVES} The cross-dispersed UV-Visual Echelle Spectrograph (UVES) is mounted at UT2-Kueyen of the VLT at Cerro Paranal \citepads{2000SPIE.4008..534D}. It was used to obtain high-resolution spectra of LB-1, HR\,2309, HR\,3195, HD\,44637, and V1371\,Tau.
The slit width was set to 1", providing a resolving power of $R\sim40\,000$. 
LB-1 was observed in the 437/760 cross-disperser setting, giving simultaneous spectral coverage in the ranges of about 380-495\,nm and 570-950\,nm. 
HR\,2309, HR\,3195, HD\,44637, and V1371\,Tau were observed in the 390/580 setting, giving coverage of the 330-450\,nm and  480-680\,nm spectral regions. 
The data were reduced with the ESO {\it Reflex} pipeline workflow for UVES \citepads{2013A&A...559A..96F}. {In all cases, including LB-1 as the faintest target, and data from the two smaller telescopes below, the signal-to-noise ratio was well above 100 over the entire spectrum.}

\subsubsection{FEROS}\label{sec:obs:feros}
The HR\,6819 spectra were taken with the {\it Fibre Extended Range Optical Spectrograph} \citep[{\sc Feros},][]{1999Msngr..95....8K} and reduced with the standard {\sc Feros}  pipeline\footnote{\url{https://www.eso.org/sci/facilities/lasilla/instruments/feros/tools/DRS.html}}. The instrument provides a coverage of about 370-920\,nm with a resolving power of $R\sim48\,000$. These are the same spectra as already used by \citetads{2020A&A...637L...3R} and introduced there in detail. In addition, this study uses two archival spectra of HR\,4930\footnote{\url{https://archive.eso.org}}, that were reduced the same way.

\subsubsection{ARCES}

The Astrophysical Research Consortium Echelle Spectrograph \citep[ARCES;][]{2003SPIE.4841.1145W}, mounted at the 3.5\,m ARC-telescope of the Apache Point Observatory, is a high resolution, cross-dispersed visible light spectrograph with a coverage of 320-1000\,nm and a resolving power of $R\sim31\,500$. ARCES observed 141\,Tau, V742\,Cas, V1362\,Cyg, V2174\,Cyg, V447\,Sct, HR\,2309, HR\,8107, and HD\,44637 for this study. The data were reduced using procedures from the Image Reduction and Analysis Facility (IRAF\footnote{IRAF is distributed by the National Optical Astronomy Observatories, which are operated by the Association of Universities for Research in Astronomy, Inc., under cooperative agreement with the National Science Foundation.}) including 1D extraction, bias subtraction, removal of scattered light and cosmic rays, division by flat field exposures, wavelength calibration via ThAr lamp exposures, as well as continuum normalization and merging of the 107 orders.

\subsection{Other}\label{sec:obs:other}
As mentioned above, candidates were as well partly identified by visual inspection of archival data, in particular the spectra in the BeSS database
\citep{2011AJ....142..149N,2018sf2a.conf..459N}. This is a large collection of Be star spectra taken by amateur astronomers worldwide over the past two decades. Spectrographs range from long-slit low resolution instruments over relatively high-resolution \mbox{{H}$\mathrm{\alpha}$}\xspace spectrographs to fiber coupled echelle instruments spanning most of the visual domain with a resolving power of typically $R\sim 8000$ to $11\,000$. The spectra of higher quality available for the identified candidates are used in this study, their technical details can be obtained from the BeSS database\footnote{\url{http://basebe.obspm.fr/basebe/}}.  Upon request, additional spectra were taken by BeSS observers to improve or clarify period values. 
Further, a few single spectra were obtained from other public archives or published resources, which are referenced in the text where applicable.

All parallaxes mentioned in this work have been taken from Gaia DR3 \citep{2023A&A...674A...1G}.
However, because the confirmed targets are all as well astrometric binaries, these parallaxes cannot be translated into distances directly, and indeed all stars mentioned below show signs of acceleration anomaly in their astrometric data \citep{2019A&A...623A..72K,2021ApJS..254...42B}. {None of the targets, however, have an entry in the binary tables of Gaia DR3 \citep{2022yCat.1357....0G}.}
For some candidates, photometric data obtained by TESS \citep[{Transiting Exoplanet Survey Satellite},][]{2015JATIS...1a4003R} has been used\footnote{\url{https://mast.stsci.edu}}. 

\section{Individual Systems}\label{sec:systems}
In this section, the individual systems are discussed, first in order of discovery (LB-1 and HR\,6819), then by confidence level, where the systems with interferometric confirmation come first (V742\,Cas, HD\,44637), then the systems with a well determined spectroscopic orbit and sufficient own data for an independent analysis, where a distant orbit could be excluded by interferometry (V447\,Sct, V1362\,Cyg), followed by systems with only some of our own data (V2174\,Cyg), and finally systems that will need further investigation before any conclusion can be drawn (HR\,8107, HR\,4930, V505\,Mon, V658\,Car) or rejected candidates (141\,Tau, HR\,2309, HR\,3195, V1371\,Tau). \mbox{{H}$\mathrm{\alpha}$}\xspace profiles and the blue and red spectral regions for most of those objects are shown in the Appendix as Figs.~\ref{fig:app:Ha}, \ref{fig:app:blue}, and \ref{fig:app:red}.

The orbits were determined using orbfit-lib\footnote{\url{https://www.chara.gsu.edu/analysis-software/orbfit-lib}}.
In all cases a free orbital solution resulted in an eccentricity with zero within the uncertainties, so that the solution was fixed to a circular orbit.

\subsection{LB-1 (LS\,V\,+22\,25, ALS\,8775)}
This is the first system that was suggested to consist of a Be star with a stripped and bloated low-mass, i.e., just post-overflow, companion that dominates the visual absorption spectrum, while the Be star is {mostly} seen by virtue of its line emission, {with only a marginal contribution due to the absorption} \citepads{2020AA...639L...6S}.  \citetads{2020Natur.580E..11A} concluded that the stripped companion is almost non-rotating, with $\mbox{$v\sin i$}\xspace=7.5$\,\mbox{${\mathrm{km\,s}}^{-1}$}\xspace and a macroturbulence of $\xi=4$\,\mbox{${\mathrm{km\,s}}^{-1}$}\xspace. A more detailed chemical analysis by \citetads{2020A&A...633L...5I} revises the rotation to $\mbox{$v\sin i$}\xspace=8.7\pm0.2$\,\mbox{${\mathrm{km\,s}}^{-1}$}\xspace with macroturbulence $\xi\leq0.1$\,\mbox{${\mathrm{km\,s}}^{-1}$}\xspace. 

The spectrum of the system is shown in the Appendix~\ref{app:spectroscopic}. It is relatively cool, in the late-B-type regime (about 13kK judging from the \ion{Si}{ii}/{\sc iii} balance, and 12720\,K according to \citealt{2020A&A...633L...5I}), and shows an obvious CNO-process-affected abundance pattern, or as stated by \citeauthor{2020A&A...633L...5I}: "He and N are strongly enriched at the expense of C and O." These authors also report a systematic under-abundance of Mg, Al, Si, S, Ar, and Fe. However, that analysis was undertaken with the understanding of the spectrum as originating from a single star. The dilution of these lines due to the continuum contribution of the Be star might explain the systematic weakness of these lines, even for solar abundances. The composite nature also explains the shape of the helium lines, for which the authors state that "observed line wings are too strong while the line cores are too weak" for a normal stellar atmosphere. The spectrum of the Be star shows most clearly in \ion{He}{i} lines, suggesting a much earlier B subtype.

The first of the two interferometric observations taken one year apart with VLTI/GRAVITY had to be discarded due to a bad calibrator, which displayed a strong binary-like signal. The second one does not show any indication of a third, distant companion, as the closure phases are consistent with zero. The bandwidth-smearing FoV of the measurements is $\sim$84\,mas, and searching for a companion up to a separation of 120\,mas with \textit{CANDID} led to a null result. The mean of the minimum magnitude difference $\Delta K$ between the primary Be star and a possible companion at separations $<120$\,mas is $4.69\pm0.25$\,mag. This rules out the black hole scenario within a triple system, as has been suggested (and as well disproven, see below) for HR\,6819. The angular resolution of the VLTI with UTs of the order of 2\,mas, together with the distance to LB-1 above 1\,kpc according to the Gaia DR3 parallax, do not enable resolving a binary system with the given orbital period.

\begin{figure*}[t]
\begin{center}
\includegraphics[width=\textwidth]{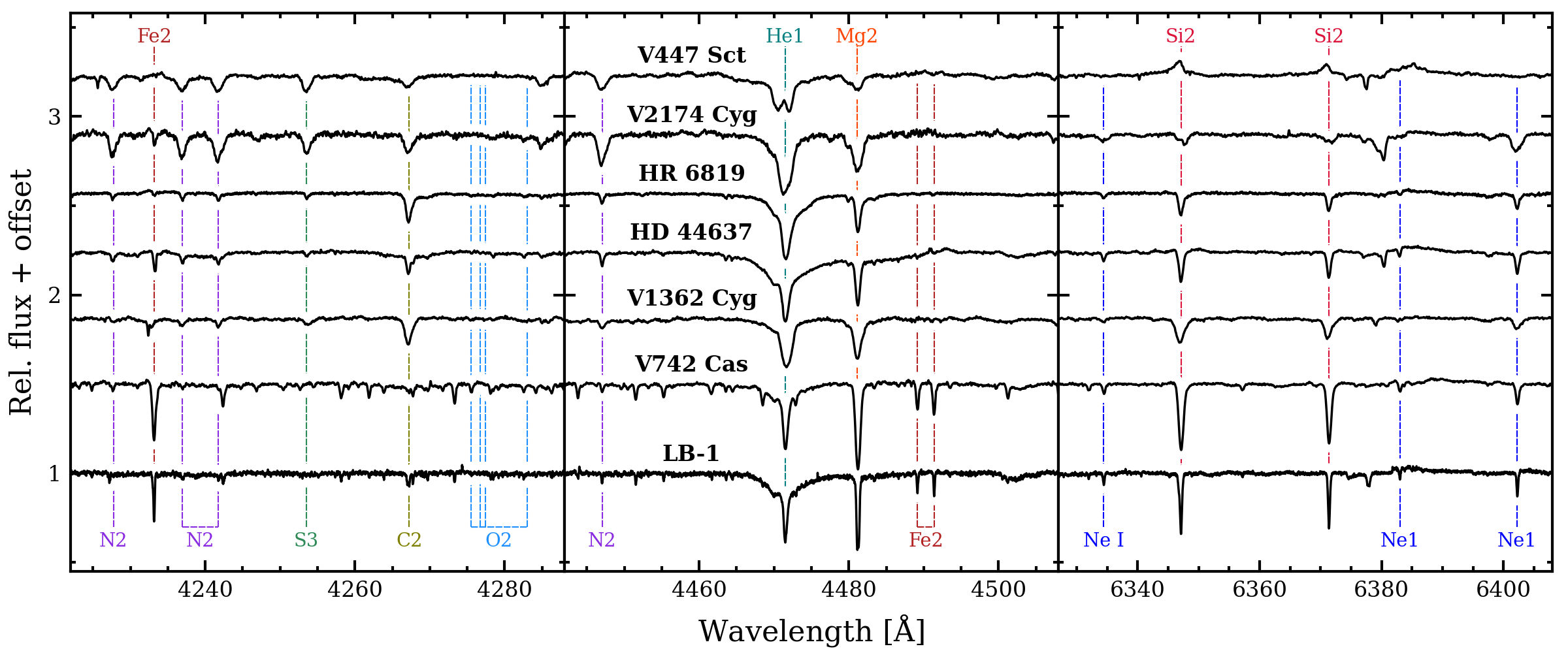}
\end{center}
\caption[xx]{\label{fig:spec}Spectral regions of several candidates with high-quality spectra showing lines of interest with respect to the chemical nucleosynthesis. From bottom to top, spectra are sorted in approximate order of increasing effective temperature. 
The presence/absence of the \spec{C}{ii}{4267} in the various spectra and the strength of nitrogen lines indicates non-solar chemistry. The RV-variable spectra have been shifted to line up their spectral features. {Except for LB-1 (UVES) and HR\,6819 (FEROS), all spectra were obtained with ARCES.}
See Figs.~\ref{fig:app:blue} and \ref{fig:app:red} for more comprehensive plots.
}

\end{figure*}

\subsection{HR\,6819 (HD\,167128, QV\,Tel)}
This is the second system for which a Be+pre-subdwarf nature was suggested, by \citet{2020AA...641A..43B}, and then proven by \citet{2022AA...659L...3F}. The former qualitatively find  $\mbox{$T_{\mathrm{eff}}$}\xspace=16\pm1$\,kK and the "indication for an abundance pattern in line with expectations for CNO-processed material" for the narrow lined component, but no significant helium overabundance. This suggests that the core has not been stripped very deep into the hydrogen burning shell. It should be noted, though, that \citeauthor{2020AA...641A..43B} explicitly only mention nitrogen and oxygen, not carbon. The latter should be depleted, but seems quite strong in the spectra (see Fig.~\ref{fig:spec}). It is also noteworthy that HR\,6819 has a rather strong \spec{Ne}{i}{6402} line, compared to \spec{Si}{ii}{6437}, yet also obvious \ion{C}{ii} lines, which may indicate a more progressed stage of nucleosynthesis than CNO burning, although this needs to be modeled in detail. 

\citet{2020AA...641A..43B} give a low rotational velocity of $\mbox{$v\sin i$}\xspace=15$\,\mbox{${\mathrm{km\,s}}^{-1}$}\xspace, albeit with a very high macroturbulence of $\xi=35$\,\mbox{${\mathrm{km\,s}}^{-1}$}\xspace, but also note that they exclude $\mbox{$v\sin i$}\xspace>25$\,\mbox{${\mathrm{km\,s}}^{-1}$}\xspace in any case, which we consider a safe upper limit. 
Further spectral properties for the pre-subdwarf are discussed by \citetads{2020A&A...637L...3R}; in particular their Figs.~2 and C.2 are of interest for comparison with similar figures of the candidate objects shown below. For the Be star, \citet{2020AA...641A..43B} estimate an effective temperature of 20\,kK, again an earlier spectral subtype than for the pre-subdwarf.

\subsection{V742\,Cas (HD\,698)}

\begin{table}[b]
\caption[xx]{\label{tab:V742CasOrbElem}Orbital elements for V742\,Cas. 
}

\begin{center}
\begin{tabular}{lcc}
\hline\hline
Parameter  & This work             & \citetads{1992ApJS...81..303S} \\
\hline
Period     & $55.9305\pm0.0034$\,d   & $55.9233\pm0.0009$\,d \\
$T_0$      & $56050.939\pm0.227$ 
                                     & $43316.585\pm0.11$ \\
$e$        & $0.0$                &  $0.005\pm0.001$  \\
$K$        & $87.06\pm1.24$  \,\mbox{${\mathrm{km\,s}}^{-1}$}\xspace        & $81.2\pm1$\,\mbox{${\mathrm{km\,s}}^{-1}$}\xspace \\
$\gamma$   & $-22.94\pm0.87$\,\mbox{${\mathrm{km\,s}}^{-1}$}\xspace & $-27.1\pm0.6$\,\mbox{${\mathrm{km\,s}}^{-1}$}\xspace \\
$i$        & $119.44\pm0.27^\circ$ \\
$a$        & $0.663\pm0.003$\,mas \\
$\Omega$   & $114.82\pm0.31^\circ$ \\ 
$\omega_{\mathrm Be}$   & $90^\circ$\\
$\omega_{\mathrm {pre-sd}}$   & $270^\circ$\\
\hline
\end{tabular}
\end{center}
\end{table}

\begin{figure*}[t]
\begin{center}
\includegraphics[width=18cm]{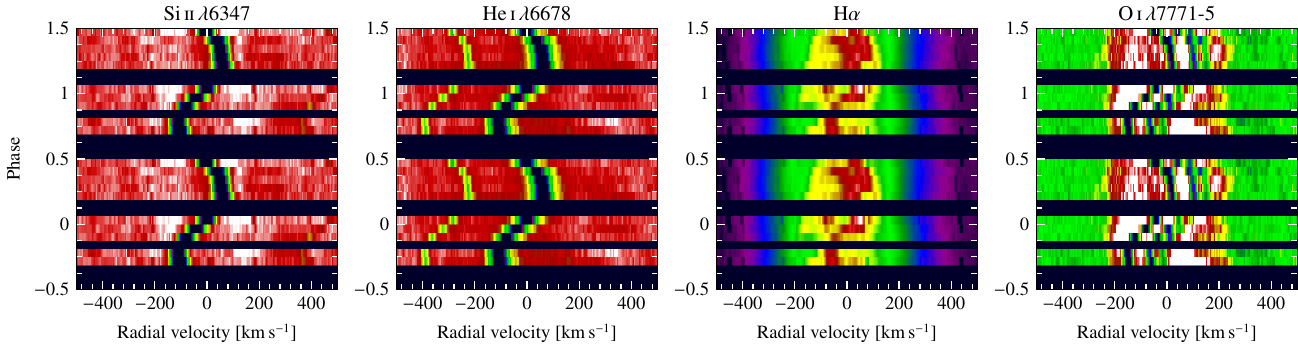}%
\end{center}
\caption[xx]{\label{fig:V742Cas_dyn}ARCES spectra of V742\,Cas folded with the orbital period for selected lines as labeled. The absorption components in the \ion{O}{i}, \ion{Si}{ii} and \ion{He}{i} lines (seen as dark trails) originate from the photosphere of the stripped star. With the exception of the RV variable peak in \mbox{{H}$\mathrm{\alpha}$}\xspace\ the emission originates from the disk regions close to the Be star. {Due the phased RV variability that single peak, however, must still be linked dynamically to the system, and may} be the signature of some still ongoing, low level mass transfer, affecting the outer disk, between the two components.}

\end{figure*}
\begin{figure*}
\begin{center}
\includegraphics[angle=0,width=8cm,clip]{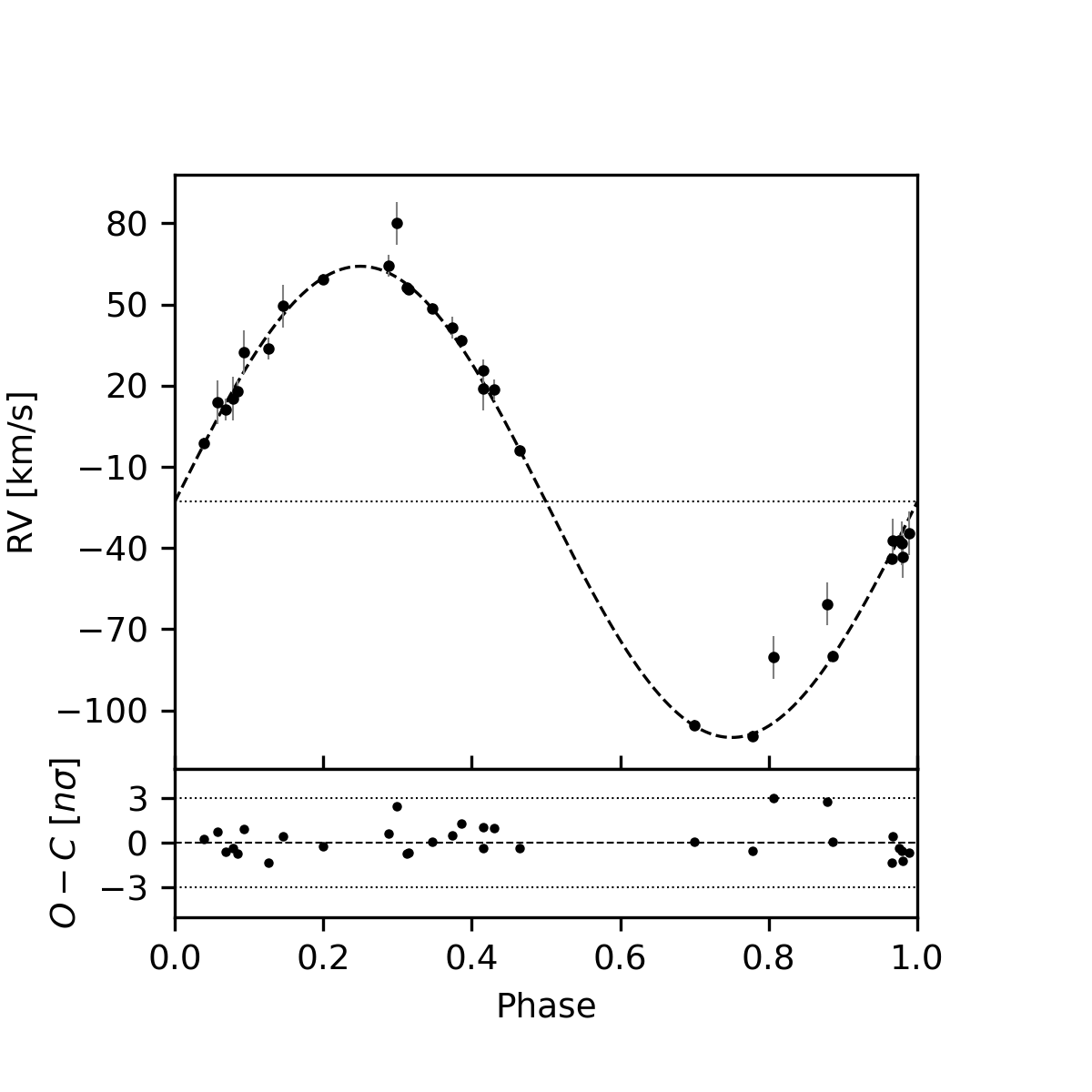}%
\includegraphics[angle=0,width=8cm,clip]{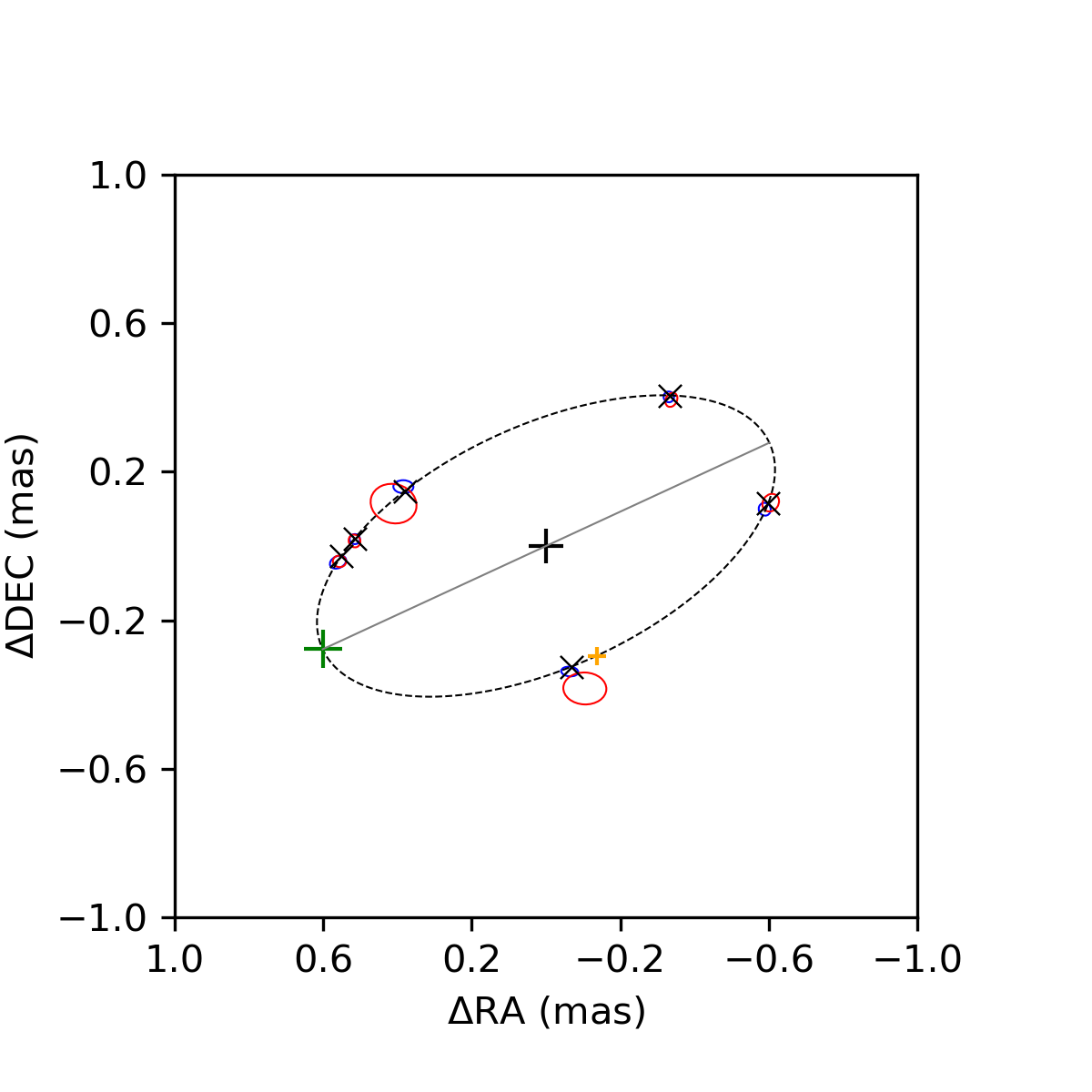}%
\end{center}
\caption[xx]{\label{fig:V742Cas} Circular orbit solution for V742\,Cas (Table~\ref{tab:V742CasOrbElem}). Left: RV curve of the narrow-lined component (dashed line) with the RVs measured in the \spec{He}{i}{6678} line (black circles). The bottom panel shows the residuals in units of $\sigma$. Right: Relative astrometric orbit derived from interferometry. The orbital motion is clockwise. The black plus sign corresponds to the location of the primary Be star, while the best-fit companion orbit is plotted as a black dashed line. Error ellipses correspond to the interferometric measurements with 5-$\sigma$ uncertainties (MIRC-X in blue and MYSTIC in red). The associated calculated positions are plotted as black crosses. Finally, the line of nodes is plotted as gray solid line, and the green cross shows the ascending node and the smaller yellow one the superior conjunction of the Be star that is used as $T_0$ for the phased plots. The two MYSTIC points with larger error ellipses correspond to angular separations below the nominal resolution of CHARA in the $K$-band.}

\end{figure*}
This system is listed by \citetads{2002ASPC..279..143C} as having an invisible companion, with reference to \citetads{1992ApJS...81..303S}, although it was known to be a single-lined binary since 1932 (\citeauthor{1932MNRAS..92..877P}). \citetads{1992ApJS...81..303S} derived the orbit from UV spectroscopy taken with the International Ultraviolet Explorer satellite, IUE, and discuss wind properties based on the typical wind lines expected in an OB-type star. Spectroscopic observations taken from the BeSS database did confirm the orbit with a period of 55.9\,d and a radial velocity amplitude $K$ of about 80\,\mbox{${\mathrm{km\,s}}^{-1}$}\xspace in a circular orbit. The system was then observed with ARCES and CHARA, which again confirmed the spectroscopic orbit, and revealed it to be a low-contrast binary, i.e., one with similar brightness of the two components.

A combined orbital solution was then obtained, using BeSS, ARCES, and CHARA observations. For the radial velocities, the \spec{He}{i}{6678} line was used (see Appendix (\ref{app:spectroscopic}). This is present in 12 amateur spectra of the H$\alpha$ region, 7 echelle spectra taken by amateur observers (all those are available from the BeSS database), and 12 ARCES spectra (see Table~\ref{tab:V742CasRV}).

The ARCES spectra show quite narrow, but well resolved deep absorption lines, with a typical full width at half maximum (FWHM) of $\sim35$\,\mbox{${\mathrm{km\,s}}^{-1}$}\xspace for lines of \ion{Fe}{ii} or \ion{Si}{ii}. This can be compared to  \citeauthor{2020AA...641A..43B}'s analysis of HR\,6819, which has a FWHM of about 60\,\mbox{${\mathrm{km\,s}}^{-1}$}\xspace.  A preliminary spectroscopic analysis, which will be published elsewhere in full once complete {(Przybilla, priv.\ comm.)}, suggests a \mbox{$v\sin i$}\xspace of 11\,\mbox{${\mathrm{km\,s}}^{-1}$}\xspace with a macroturbulence $\xi$ of 19\,\mbox{${\mathrm{km\,s}}^{-1}$}\xspace.

{The spectrum does not show solar abundances, but is He-rich, with nitrogen enhanced and carbon depleted (Przybilla, priv.\ comm.). As a quantitative analysis is reserved for a future work}, it is difficult to estimate the photospheric temperature from the usual indicators.  But it should be noted that lines of both \ion{Fe}{iii} and \ion{Si}{iii} are present, though much weaker than those of \ion{Fe}{ii} and \ion{Si}{ii}, and that the spectrum is generally similar to that of LB-1 (which has $\sim$13\,kK), including the chemical peculiarities, apart from the broader lines (see Fig.~\ref{fig:V742Cas_dyn} and Appendix \ref{app:spectroscopic}).  The Be-typical emission lines are stable in radial velocity, although their general weakness other than in \mbox{{H}$\mathrm{\alpha}$}\xspace makes it difficult to give an upper limit to any RV variability that would mirror that of the narrow-line component. {Attempts to measure the \mbox{{H}$\mathrm{\alpha}$} radial velocity variations through spectral disentangling were unsuccessful; neither were any traces of the Be star absorption spectrum found.}

The interferometric results are summarized in Table~\ref{tab:V742Cas_astrometry}. From the four measurements with separations $>0.5$\,mas, the fainter component contributes on average $31.1 \pm 0.3$\% of the total flux in the $H$ band and $35.5 \pm 0.4$\% of the total flux in the $K$ band. 
However, with the data at hand it is not possible to distinguish whether the Be star or the narrow-lined component is the brighter one, since the Be star cannot be identified with the low spectral resolution of MIRC-X or MYSTIC. The fact that the fainter star seems to be brighter in $K$ than in $H$, i.e., it is redder, would suggest that the Be star is the fainter component, because the disk contribution to the Be-star light reddens the spectral energy distribution (SED). 

The combined solution, i.e. taking both spectroscopy and astrometry into account and presented in Fig.~\ref{fig:V742Cas}, agrees well with the IUE-based solution by \citet[][see Table~\ref{tab:V742CasOrbElem}]{1992ApJS...81..303S}. Yet, this solution translates to unphysical parameters when the Gaia DR3 distance of 708\,pc is used, such as a negative mass for one component. However, that is not necessarily a reliable distance, since even though it has a Renormalised Unit Weight Error (RUWE) of only 1.46, it is listed as a star with a proper motion anomaly by \citetads{2019A&A...623A..72K} and \citetads{2021ApJS..254...42B}. With a semi-major axis of 0.663\,mas this is the smallest visually confirmed orbit to date, with the previous record-holder being \object{WR\,133} \citepads{2021ApJ...908L...3R}. Finally, as a binary, it should be removed from the interferometric calibrator catalog by \citetads{2019MNRAS.490.3158C}.  

A more detailed analysis of the orbital parameters, spectral disentangling of the components, the distance, the component masses, and the physical properties including the abundances is deferred to a later study. The preliminary analysis {done for that study by Przybilla (priv.\ comm.)} does confirm the basic result that, despite the about unity light ratio, one component is of a normal mass for an intermediate B star and has largely solar composition, while the other has less than one fifth of that mass and shows CNO-cycle-modified chemistry.


\subsection{HD\,44637}
HD\,44637 was identified in the BeSS database as an object with narrow and RV-variable spectral absorption lines, while the emission did not show the same behavior.  It was observed a few times with ARCES and UVES, and in a coordinated campaign through BeSS. The absorption line radial velocity (RV) is variable, but the period proved to be too long to obtain a solid value before the end of the observability period 2022/23. It is certainly longer than half a year.

Two GRAVITY/VLTI observations show an unambiguous detection of two unresolved sources separated by about 1.5\,mas, with a contrast ratio of about 40\% to 60\%, where the Be star could be identified as the brighter source by virtue of the phase signature across the Br$\gamma$ line.  The two observations were taken about two months apart, and the position angle between the two components changed by about 90 degrees. This is in agreement with RV measurements obtained from BeSS spectra, that show a slow redward acceleration by about 25\,\mbox{${\mathrm{km\,s}}^{-1}$}\xspace over the same time span. This also indicates a long period of at least half a year or more. As seen later in the comparison of all systems, this is unusually long, but not unique.
\mbox{{H}$\mathrm{\alpha}$}\xspace shows the strongest emission of the sample, at an emission height in units of the continuum of $E/C=8$ in the combined spectrum, so for the Be star alone it must be well above 10 (Fig.~\ref{fig:app:Ha}), and Br$\gamma$ still $E/C=3$ (for the Be star alone, see Fig.~\ref{fig:app:HD44637_PMOIRED}). Such a well developed disk is another indication for a long period, since the radius of tidal truncation of the disk must be accordingly large.

The stellar spectrum is compared best to HR\,6819, as, judging from the \ion{Si}{ii}/\ion{Si}{iii} ratio it is only slightly cooler, and has a similar rotational velocity \mbox{$v\sin i$}\xspace. It also has enhanced strength of the nitrogen lines and compared to HR\,6819 even less carbon and less oxygen, but even though it is cooler the helium lines are stronger. 

Although a full determination of the orbital elements is ongoing due to the long period, and results must be deferred to a later study, this is not needed to count HD\,44637 as a binary of the same nature as HR\,6819, and with the same confidence, as the interferometric and spectroscopic variability is fully clear in that respect.

\subsection{V447\,Sct (HD\,173219)}

This system is also listed by \citetads{2002ASPC..279..143C} as a B-type star with invisible companion.  While it is sometimes classified as a B0 supergiant, this is probably not the case. Its parallax is 0.5\,mas. For comparison, the actual B1 supergiant $\zeta^1$\,Sco has the same parallax value, but is 3.5\,magnitudes brighter in $V$. Unless there is highly unusual extinction towards V447\,Sct, its magnitude is more in line with a core hydrogen burning early B star, and even more so if it is a binary with similarly bright components, i.e., low contrast. This is also reflected by some earlier classifications as B1:V:npe \citep{1955ApJS....2...41M,1956ApJS....2..389H}.  The spectrum was analyzed in more detail by \citetads{2006A&A...451.1053F}, who confirm a more classical Be-like spectral type of B0.5\,IV, and also give $\mbox{$v\sin i$}\xspace=61$\,\mbox{${\mathrm{km\,s}}^{-1}$}\xspace for the absorption lines, which, however, do not belong to the Be star in the system.

\begin{figure}[t]
\begin{center}
\includegraphics[angle=0,width=8cm,clip]{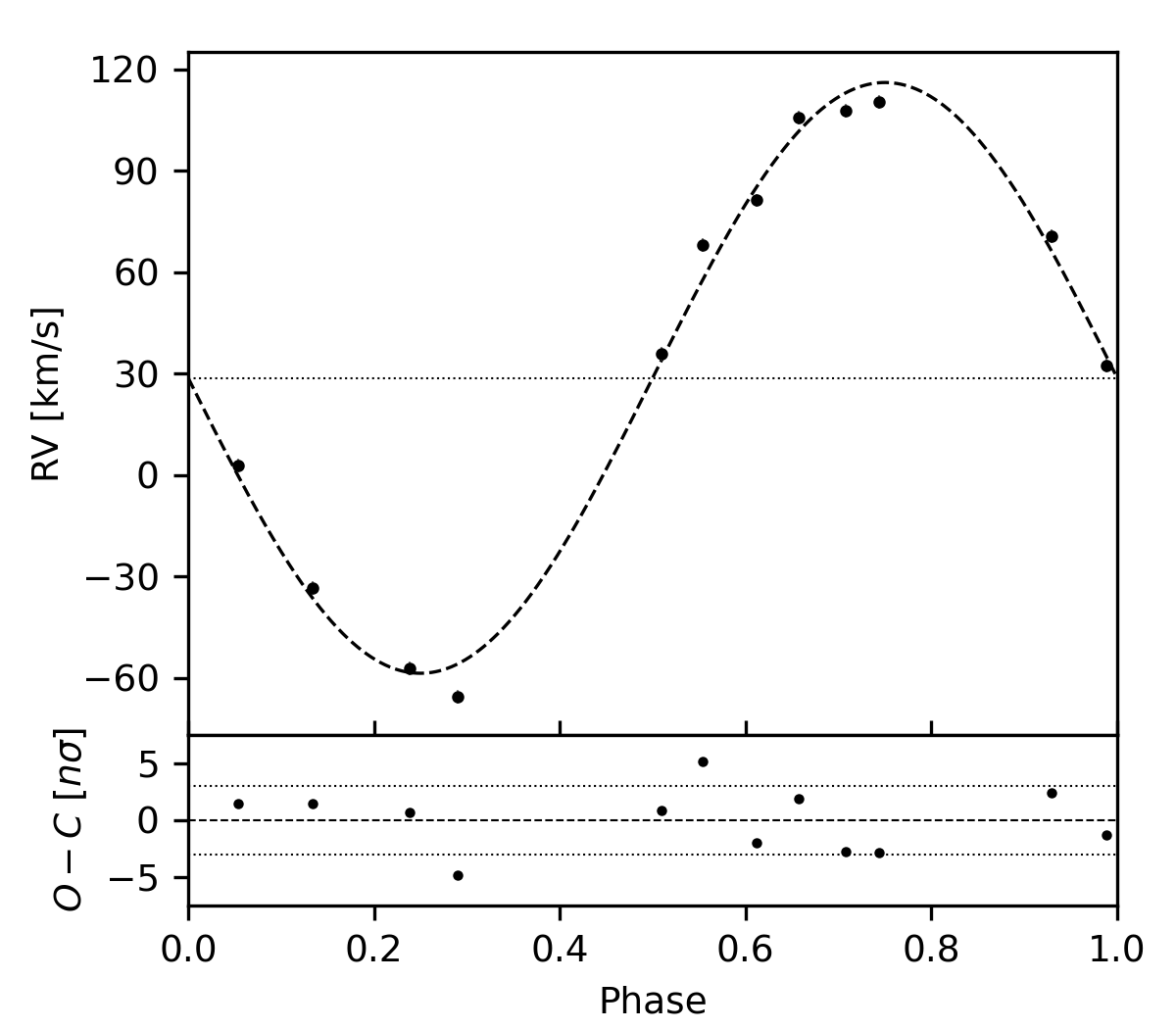}

\end{center}
\caption[xx]{\label{fig:V447Sct} Same as the left panel of Fig.~\ref{fig:V742Cas}, but for a circular orbit solution for V447\,Sct (Table~\ref{tab:V447SctOrbElem}), and RVs measured in the \spec{Si}{iii}{4553} line.
}

\end{figure}

\begin{figure*}[t]
\begin{center}
\includegraphics[width=18cm]{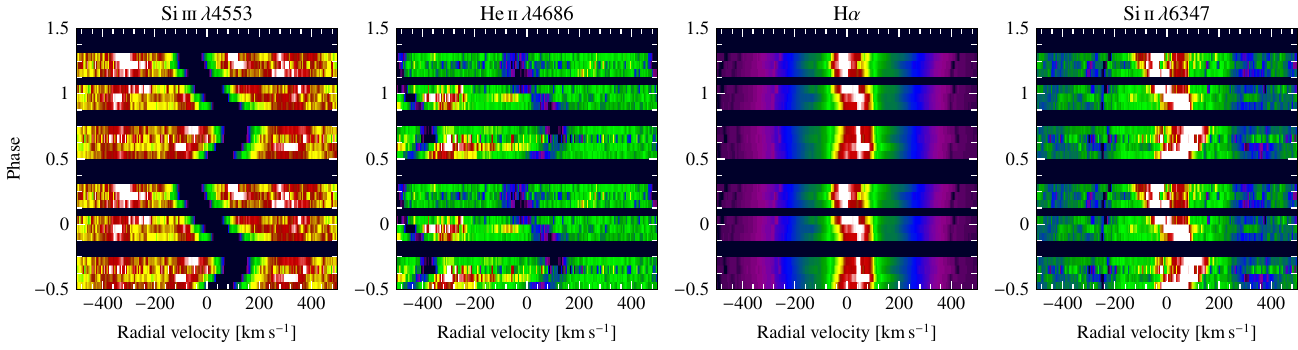}%
\end{center}
\caption[xx]{\label{fig:V447Sct_dyn}ARCES spectra of V447\,Sct folded with the orbital period for selected lines as labeled. Similar to {V742\,Cas shown in} Fig.~\ref{fig:V742Cas_dyn}, except for the single peaked emission in \ion{Si}{ii}, something which, however, is also seen in HR\,6819 \citepads{2020A&A...637L...3R}. }

\end{figure*}

\citetads{1973MNRAS.163..219H} identify the system as very peculiar, and, as with other systems mentioned here, having a very massive, invisible secondary, which they speculate could be a collapsed object like a black hole.
Their spectroscopic orbit as well puts it squarely in the same regime as the other candidates, with $ P=58.41$\,d, $K=87.4$\,\mbox{${\mathrm{km\,s}}^{-1}$}\xspace and a circular orbit (see Figs.~\ref{fig:V447Sct} and \ref{fig:V447Sct_dyn}). The FWHM of the absorption lines is quite high, in agreement with the \mbox{$v\sin i$}\xspace value of \citetads{2006A&A...451.1053F}, but the profiles are Gaussian shaped, similar to those of HR\,6819, so that macroturbulence might play a significant role in their broadening, not just rotation.

The H$\alpha$ emission is {morphologically} unique. It is certainly not a P\,Cygni wind type profile but neither that of a classical Be star. It has a broad base, that does not vary in radial velocity with any obviously {seen} amplitude in the given spectra. On top of that base there is a single peak that does move in RV with the binary period (see Fig.~\ref{fig:app:Ha}). This is similar to what is seen in V742\,Cas, but here the {RV-amplitude} of the H$\alpha$ emission peak is only half that of the photospheric absorption lines, indicating either some active accretion stream still present in the system, or a wind-wind interaction. In lines other than \mbox{{H}$\mathrm{\alpha}$}\xspace with single-peaked emission, however, the amplitude is comparable to that of the absorption lines. These include \spec{He}{i}{5876}, \spec{He}{i}{7061}, {\spec{He}{i}{4471} (where it mimics a double absorption line),} and \spec{Si}{ii}{6347} (Fig.~\ref{fig:V447Sct_dyn}).

The absorption spectrum is the earliest among the candidates, showing strong \ion{Si}{iii} lines and a \spec{He}{i}{4471}/\spec{Mg}{ii}{4481} balance indicative of an early-type Be star. {The spectrum has very obvious strong nitrogen and helium lines, but nearly absent carbon and oxygen lines. This strongly suggests CNO-processed material.}

V447\,Sct was observed with CHARA, but this did not result in the detection of a companion. This is not too unexpected because V447\,Sct is fainter and farther away than V742\,Cas. Two observations with CHARA were taken 14 days apart close to phases 
$\sim$0.32 and $\sim$0.55 according to the ephemeris given in Table \ref{tab:V447SctOrbElem}, at least partly
corresponding to epochs close to RV maxima. 
No companion was revealed with a minimum magnitude difference $\Delta H_{\mathrm{min}} = 3.70\pm0.16$\,mag and $\Delta K_{\mathrm{min}} = 3.35\pm0.19$\,mag at the first epoch and $\Delta H_{\mathrm{min}} = 2.75\pm0.22$\,mag and $\Delta K_{\mathrm{min}} = 4.11\pm0.17$\,mag at the second epoch for angular separations between 0.5 and 25\,mas.

\begin{table}[b]
\caption[xx]{\label{tab:V447SctOrbElem}Orbital elements for V447\,Sct, derived from the RVs listed in Table~\ref{tab:V447SctRV}. 
}
\begin{center}
\begin{tabular}{lc}
\hline\hline
Period     & $58.41\pm0.06$\,d   \\
$T_0$      & $57992.3\pm0.4$     \\
$e$        & $0.0$       \\
$K$        & $87.4\pm2.9$\,\mbox{${\mathrm{km\,s}}^{-1}$}\xspace \\
$\gamma$   & $28.7\pm1.9$\,\mbox{${\mathrm{km\,s}}^{-1}$}\xspace \\
$\omega$   & $90^\circ$   \\
\hline
\end{tabular}
\end{center}
\end{table}

\subsection{V1362\,Cyg (HD\,190467)}

\begin{figure}[t]
\begin{center}
\includegraphics[angle=0,width=8cm,clip]{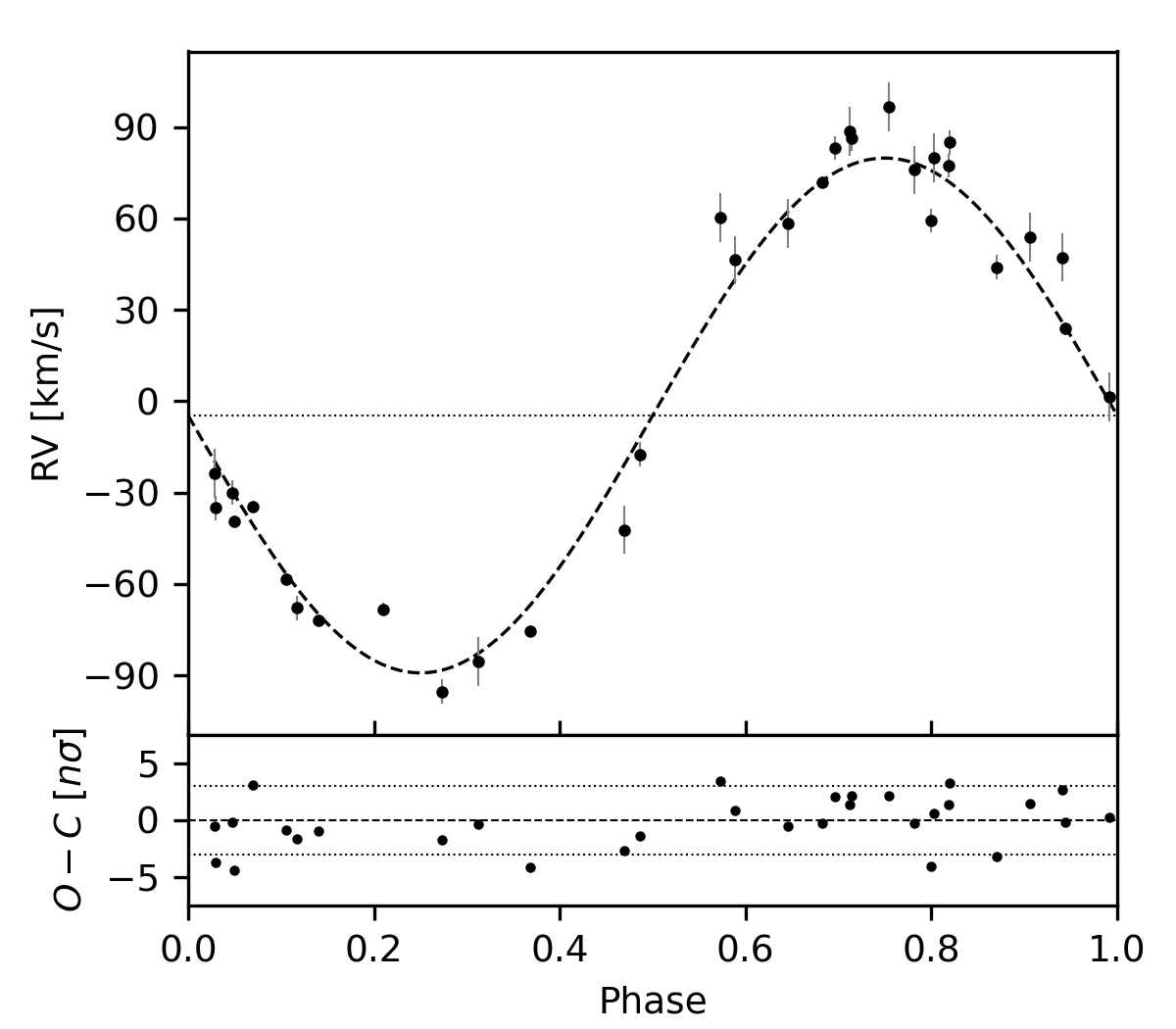}
\end{center}
\caption[xx]{\label{fig:1362Cyg} Same as Fig.~\ref{fig:V447Sct}, but for a circular orbit solution for V1362\,Cyg (Table~\ref{tab:V1362CygOrbElem}), and RVs measured in the \spec{He}{i}{6678} line.
}

\end{figure}

\begin{table}[b]
\caption[xx]{\label{tab:V1362CygOrbElem}Orbital elements for V1362\,Cyg, derived from the RVs listed in Table~\ref{tab:V1362CygRV}.
}
\begin{center}
\begin{tabular}{lc}
\hline\hline
Period     & $56.847\pm0.028$\,d   \\
$T_0$      & $57485.63\pm0.94$     \\
$e$        & $0.0$       \\
$K$        & $84.50\pm2.60$\,\mbox{${\mathrm{km\,s}}^{-1}$}\xspace \\
$\gamma$   & $-4.7\pm2.0$\,\mbox{${\mathrm{km\,s}}^{-1}$}\xspace \\
$\omega$   & $90^\circ$   \\
\hline
\end{tabular}
\end{center}
\end{table}
This system, too, is listed by \citetads{2002ASPC..279..143C} as high-mass-function and invisible-companion binary.  The source given there, \citepads{1976PDAO...15....1H}, only refers to an unpublished orbit with $\sim57$\,d, but states it has been "thoroughly checked" to rule out other periods. There are 24 spectra in BeSS taken over more than 10 years suitable for RV measurement of \spec{He}{i}{6678}, as well as a number of recent ARCES spectra, see Table~\ref{tab:V1362CygRV}.  From them, the orbit was determined independently, confirming a circular orbit with a period of 56.81\,d and an RV amplitude of $K=85$\,kms for the absorption lines (Fig.~\ref{fig:1362Cyg}). This translates to an orbital radius of about 0.5\,au$/\sin i$. Considering the parallax of less than 1\,mas the interferometric non-detection is unsurprising: CHARA/MIRC-X measurement in the $H$ band revealed no low contrast companion and a mean $\Delta  H_{\mathrm{min}}$ of $3.19\pm0.12$\,mag for angular separation between 0.5 and 25\,mas. It should be kept in mind, though, that the observing date was close to phase zero according to Table~\ref{tab:V1362CygOrbElem}, so this is not a fully conclusive non-detection, depending on inclination.

The classification as eclipsing binary present in catalogs \citep{1984AISAO..18...64P,1985ApJ...295..143M,2006A&A...446..785M,2013AN....334..860A}, with a period of 7\,d, can be traced back to \citetads{1970JRASC..64..218P}. In the one available TESS sector, a variability timescale of about 3.5\,d is present, but with semi-regular characteristics, and unlike any signature of binarity.  From a modern perspective, judging by the original plots and the TESS data, the variations are similar to the typical Be-type photometric variability that arises in the inner region of the disk as material is ejected from the star \citepads{2022AJ....163..226L}.  Therefore, V1362\,Cyg should be removed from the respective catalogs.  Likewise, being a binary, it should be deleted from the catalog of potential interferometric calibrators \citepads{2019MNRAS.490.3158C}.

\citetads{1984JRASC..78..241P} mentions a \mbox{$v\sin i$}\xspace of 75\,\mbox{${\mathrm{km\,s}}^{-1}$}\xspace, but also notes a "variable sharpness" of the lines. The typically measured FWHM in both the BeSS and ARCES spectra is about 55\,\mbox{${\mathrm{km\,s}}^{-1}$}\xspace, which is slightly narrower than the FWHM of HR\,6819 of 60\,\mbox{${\mathrm{km\,s}}^{-1}$}\xspace. For Table~\ref{tab:elem}, that summarizes the properties of the identified systems, we hence assume that it has a similar \mbox{$v\sin i$}\xspace, i.e., less than 25\,\mbox{${\mathrm{km\,s}}^{-1}$}\xspace.
The spectrum is not obviously non-solar in its abundances, since \ion{C}{ii} is quite strong, while \ion{Ne}{i} is present, but not clearly over-abundant. However, \ion{N}{ii} might have stronger lines than one would expect of a star with this temperature.

\begin{figure*}[t]
\begin{center}
\includegraphics[width=18cm]{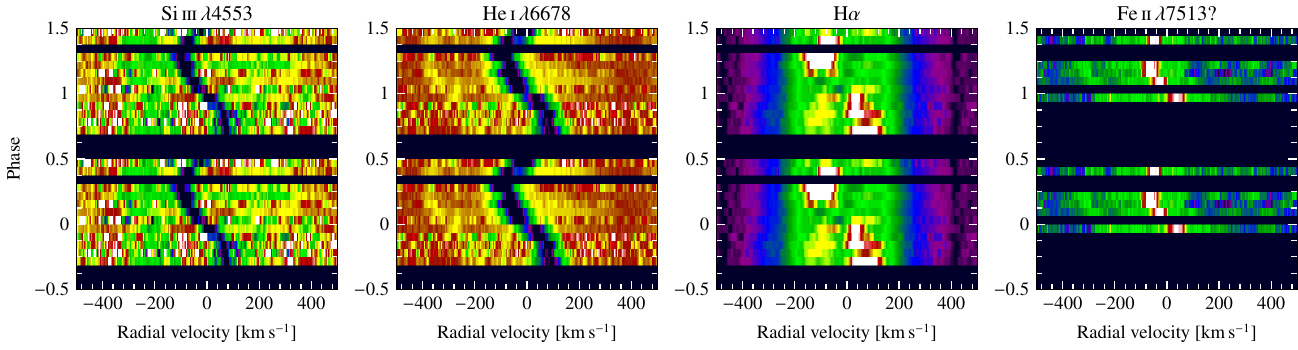}%
\end{center}
\caption[xx]{\label{fig:V1362Cyg_dyn}BeSS echelle and ARCES spectra of V1362\,Cyg folded with the orbital period for selected lines as labeled. The infrared region at 7500\,\AA\ is only covered by ARCES. Similar to Fig.~\ref{fig:V447Sct_dyn}.}
\end{figure*}

The Be-typical emission lines of V1362\,Cyg seem stationary. There are, however, two exceptions, as seen in Fig.~\ref{fig:V1362Cyg_dyn}. The first is a single, weak emission peak at a wavelength of 7513\,\AA, that traces the absorption line RV curve. This is similar to what is seen in HR\,6819 \citep[][their Fig. C.2]{2020A&A...637L...3R}. Other than in HR\,6819, however, such a single peak component is also apparent in the H$\alpha$ line, again following the absorption RV.



\subsection{V2174\,Cyg (HD\,235679)}
This is also one of the binaries noted by \citetads{2002ASPC..279..143C} to have an invisible massive companion. Spectra secured through BeSS and with ARCES corroborate the literature description as nitrogen-rich.
The identification of this star as a system with having a post-mass-overflow component is strongly suggested already by the literature. It was first noted by \citetads{1970ApJ...161..477A} as a binary, though their period of 111\,d was later rejected (see below).  \citetads{1971ApJ...164L..67W} then found it to be a nitrogen-rich B-type supergiant with strongly depleted carbon and oxygen, and suggested that it might be a new type of helium star since "it appears that the helium-to-hydrogen ratio is abnormally large". \citetads{1978ApJ...222..234B} confirmed the hydrogen deficiency and added that "there is strong emission in the hydrogen lines and weaker emission in the \ion{He}{i} lines". They derived a period of  225.16\,d, an RV amplitude of $K=63.6$\,kms, and a low eccentricity compatible with zero within $e=0.1\pm0.1$, implying a very high mass function of 5.9.   

Finally, \citetads{2001PAICz..89...23B} investigated the system in some detail and revised the period to 225.33\,d. They could not find any trace of a potentially Roche-lobe-filling companion, and consequently considered the system to consist of a Be-type supergiant and an invisible companion. However, they also state that the Be star, supposed to be a supergiant from previous works, is "subluminous for its spectral type". They largely confirm the earlier determined orbital parameters, with the exception of elements computed from hydrogen-line measurements, which they characterize as "very different from those derived from the other line groups".  They even say that the emission lines, which are strong with a height above the continuum of $E/C\sim 3$ to 4, "are approximately stationary".  They consider a black hole as the companion, but reject it in favor of a model where the emission stems from a wind-wind collision. {They also state that the spectroscopically invisible component would, then, either have to be a rapid rotator, or at least two magnitudes fainter, and even} suggest a post-overflow companion. They do not, however, consider the possibility that it is this post-overflow companion that produces the absorption spectrum they observed.

Three echelle spectra were taken with ARCES and an additional one was made available in the BeSS database. They show a largely identical spectral pattern to V447\,Sct, namely a strongly enhanced nitrogen abundance and helium enrichment, while carbon and oxygen are depleted. The \ion{Si}{ii}/\ion{Si}{iii} ionization balance points to a slightly lower temperature than that of V447\,Sct, but the red \ion{Si}{ii} lines at 6347/71\,\AA\ are probably filled in by emission. The line width is similar to that of V447\,Sct.

The findings for LB-1 and HR\,6819, the description of the other objects above, and the available spectra, leave little doubt that V2174\,Cyg is perfectly described by a model consisting of a classical, recently spun-up Be star, which is responsible for the line emission and the broad hydrogen line wings, and a low-gravity, stripped and bloated core of a post-overflow object, which contributes the narrower photospheric absorption lines and exhibits strong signs of a processed chemistry. However, at $V=9$\,mag and a distance of above 1\,kpc \citepads{2001PAICz..89...23B}, the system is very much at limit of the current interferometric capabilities for astrometric confirmation, although with knowledge of the orbit it might just be possible to resolve the system at quadrature.  For the orbital elements in Table~\ref{tab:elem}, we adopt those derived from \ion{N}{ii} by \citeauthor{2001PAICz..89...23B}, as they have the smallest uncertainties.

\begin{figure}[t]
\begin{center}
\includegraphics[angle=0,width=8cm,clip]{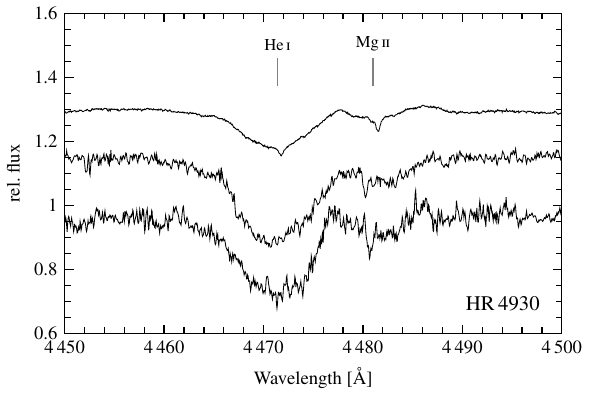}%
\end{center}
\caption[xx]{\label{fig:LSMus}The \spec{He}{i}{4471}/\spec{Mg}{ii}{4481} region of HR\,4930, clearly showing the composite nature of the spectrum. In the FEROS spectrum (upper one) also the secondary component of the \ion{He}{i} line is seen. The difference in strength of the secondary features in the spectra of \citetads{2001A&A...378..861C} with respect to the FEROS spectrum is present in the original data and remains unexplained here. }
\end{figure}


\begin{table*}
\caption{Summary of orbital elements, rotation and distance indicators of Be star binary systems hosting a pre-subdwarf companion. 
\label{tab:elem}}
\small\begin{center}
\begin{tabular}{l c c c cc cccl}
\hline\hline
 System     & Sp. type     & Period   & Amplitude      & Eccentricity      &Inclination & Rotation & Parallax& $V$-band& Sources\\
            & lit.        & $P$\,[d] & $K$\,[\mbox{${\mathrm{km\,s}}^{-1}$}\xspace]    & $e$               & $i$ [deg] & [\mbox{${\mathrm{km\,s}}^{-1}$}\xspace]  & [mas] & [mag] & for orbit\\
 \hline
 {LB-1}     &  B3V   & $78.7999\pm0.0097$ & $52.94\pm0.13$  & $0^*$ &  N/A & $7\pm2$ & 0.4 & 11.5 &1\\
{HR\,6819}   & B2V   & $40.3315\pm0.0003$  & $62.13\pm0.04$ & $0^*$ &  $49\pm1.9$ & $<25$ & 2.7 & 5.4 & 2,3\\
{V742\,Cas}  & B5II  & $55.930\pm0.004$ & $86.9\pm1.1$ & $0^*$ &  $119.4\pm0.3$ & $\sim10$ &1.4 & 7.1 &  this work \\
{HD\,44637}  & B2V:pe & $>180$ & $>30$ & N/A & N/A &   $<25$  &1.1 & 9.1 &  this work \\
{V447\,Sct}  & B1:V:npe & $58.41\pm0.06$ & $87.4\pm2.9$ & $0^*$ &   N/A  & $\sim60$ & 0.6 & 7.9 & this work \\
{V1362\,Cyg} & B5II:n  & $56.82\pm0.03$ & $82.0\pm2.9$ & $0^*$ &    N/A    &   $<25$& 0.9 & 8.2 & this work\\
{V2174\,Cyg} & B2II  & $225.33^*$  &  $61.1\pm1.2$  &  $0.075\pm0.020$ &  N/A  &     $< 60$    & 0.6 & 8.9  &  4\\
\hline
\end{tabular}
\end{center}
{\footnotesize Values without uncertainties were fixed (marked with an asterisk, $^*$). The literature spectral type in all cases refers to the pre-subdwarf absorption spectrum, not the Be star. Notes: 1: \citetads{2020AA...639L...6S}, 2: \citetads{2022AA...659L...3F}, 3: \citetads{2020AA...641A..43B}, 4: \citetads{2001PAICz..89...23B}}
\end{table*}

\subsection{Other potential candidates}


\subsubsection{HR\,8107 (HD\,201836)}

\object{HR\,8107} was discovered to be a Be star by \citetads{1986PASJ...38..627H}, so it was not part of the often meticulous observing campaigns on bright Be stars from the first half of the 20th century.  It shows a moderately broad-lined B5-type spectrum ($\mbox{$v\sin i$}\xspace \approx 160\,\mbox{${\mathrm{km\,s}}^{-1}$}\xspace$), superimposed with a much narrower and maybe slightly later B-type spectrum ($\mbox{$v\sin i$}\xspace \approx 15\,\mbox{${\mathrm{km\,s}}^{-1}$}\xspace$). There are  six ARCES spectra, three BeSS echelle spectra and a number of BeSS H$\alpha$ spectra, some of which also cover \spec{He}{i}{6678}. The narrow-lined component is clearly RV-variable, but the amplitude cannot be more than $K\sim20$\,\mbox{${\mathrm{km\,s}}^{-1}$}\xspace. The Be emission-line shape of \mbox{{H}$\mathrm{\alpha}$}\xspace suggests a low inclination, but not fully polar. \mbox{{H}$\mathrm{\alpha}$}\xspace also shows a tidally disturbed appearance in some spectra, as well as the infrared \ion{Ca}{ii} triplet in emission, both of which are hints towards a Be star with a nearby binary companion \citepads{2018A&A...609A.108S}, and not a hierarchical system of more stars, in which the Be star would be an isolated companion. In the above sample, for instance, \ion{Ca}{ii} emission is obviously present in five out of seven (LB-1, V742\,Cas, HD\,44637, V447\,Sct, and V2174\,Cyg), which is well more than the expected one-fifth from the general statistics among Be stars (\citeauthor{2018A&A...609A.108S}).

In TESS data, the object appears as an eclipsing binary with a period of 12.14569\,d \citepads{2021A&A...652A.120I}. The spectroscopic data, though, cannot be folded with that period to a coherent curve. The eclipses point to an eccentric binary of two stars of quite different radii and intrinsic brightness, but even the primary eclipse is just about 1.5\% deep. Together with the U-shape of the dip, this suggests that neither the Be star, its disk, nor the narrow lined component are likely to be involved in the eclipse.
Clues as to whether HR\,8107 is a hierarchical system or, considering the size of the TESS point spread function of about 21 arcseconds, whether the eclipsing signal comes from an unrelated background object, and how the photometric signal is related to the spectroscopic properties, require further observations, and in particular interferometry might be worthwhile.

Disregarding the eclipsing light curve, however, HR\,8107 is a quite strong potential candidate for a more pole-on case of a post-interaction system. In order to make any further statement the period and amplitude need to be established, and ideally as well an astrometric orbit, which should be easily accessible at a Gaia parallax of 1.4\,mas.

\subsubsection{HR\,4930 (HD\,113120, LS\,Mus)}\label{sec:systems:HR4930}

In the ESO archive (see Sect.~\ref{sec:obs:feros}), there are two FEROS spectra of that star, taken in direct consecution at MJD=53128.1. The spectra show a clear composite spectrum, with both objects being B-type stars, one broad and one narrow-lined (see Fig.~\ref{fig:LSMus}, uppermost spectrum). Judging by the \spec{He}{i}{4471}/\spec{Mg}{ii}{4481} ratio, as well as the presence of other species, the broad-lined component is a B star around spectral type B2, while the narrow-lined one is of type B8 or later. {It seems unlikely that the 45\,kK template used by \citet{2018ApJ...853..156W,2021AJ....161..248W} would produce a signal for any B8-type object, in particular considering the UV-weakness of such a spectral type vs.\ the suggested 4\% flux contribution.}

Narrow components are also clearly present in \ion{Si}{ii}, \ion{C}{ii}, and \ion{Fe}{ii} lines. Their typical width is about $\sim25$\,\mbox{${\mathrm{km\,s}}^{-1}$}\xspace, which fits the upper limit by \citet{2021AJ....161..248W}. The depression of the narrow \ion{Mg}{ii} feature below the ambient spectrum is about 4\%, while that of the \spec{Si}{ii}{4128/32} doublet is about 3\%. If these components had a depth of 100\% in the spectrum of the narrow-lined star, these percentages are lower limits on the fractional flux contribution of that star.   \citetads{2001A&A...378..861C} obtained two spectra of the \spec{He}{i}{4478}/\spec{Mg}{ii}{4481} region in HR\,4930.  They are of a lower quality than the FEROS data, but sufficient to identify the composite nature of \spec{Mg}{ii}{4481} (see Fig.~\ref{fig:LSMus}, lower two spectra).  These observations were obtained at MJD=47929.2 and 48641.3, respectively, and show a clear difference between the positions of the secondary \spec{Mg}{ii}{4481} component, implying a radial-velocity amplitude in \spec{Mg}{ii}{4481} of  $K>40$\,\mbox{${\mathrm{km\,s}}^{-1}$}\xspace. Unfortunately, the period  by \citet{2023AJ....165..203W} is not sufficiently precise to phase these measurements, taken about two decades and longer ago, with their SB2 solution.

There is a $\sim$3 magnitudes fainter component at a distance of 0.6", also noticed by \citetads{2021AJ....161..248W}.  However, \citet{2021AJ....161..248W} exclude that this could be the source of the subdwarf spectrum, based on the spatial resolution of HST/STIS. 

This leaves basically one option for HR\,4930: It could be a highly complex and hierarchical system, which would be the only possibility to reconcile the UV observations by \citet{2021AJ....161..248W} with the observations at optical wavelengths described above.  In such a system, components Aa and Ab would be the Be star and a subdwarf, respectively, and Ba is the B8 or later-type star with an unseen companion Bb. In order to test this hypothesis, the orbit of visual component B has to be determined, either spectroscopically or interferometrically, and compared to the observations of the UV component by \citet{2023AJ....165..203W}. If the orbit is different, it is a hierarchical quadruple system. If, however, the orbit is the same, the companion cannot be a subdwarf of 45\,kK, since such an object would have none of the singly ionized metal lines seen at optical wavelengths. Whether, in that case, the narrow-lined companion can be a normal late-B main-sequence star or would be a mass-transfer remnant would depend on the flux ratio that needs to be measured interferometrically.

\subsubsection{V505\,Mon (HD\,48914)}\label{sec:systems:V505Mon}
Two of the stars mentioned by \citetads{2002ASPC..279..143C} are not in the above list of candidates. \object{FY\,Vel} is described as a $\beta$\,Lyr system, with $P=33$\,d and $K=130$ to 140\,\mbox{${\mathrm{km\,s}}^{-1}$}\xspace \citepads{1971MNRAS.154..103T}. It probably does not fit here, but should be understood in the context of more similar systems, such as \object{$\beta$\,Lyr} or \object{V453\,Sco}. The other one, V505\,Mon, is probably more interesting in this context, and may be another candidate. However, from the literature alone this is much more difficult to assess than for V2174\,Cyg, and there are no spectra of V505\,Mon available to us. \citetads{2001A&A...375..434M} identify a gaseous disk in the system and mention the clear similarity of this system to V742\,Cas, V447\,Sct, and V1362\,Cyg: It has an orbital period of 53.8\,d and an RV semi-amplitude of $K=92.1$\,\mbox{${\mathrm{km\,s}}^{-1}$}\xspace for the absorption lines, which, at $\mbox{$v\sin i$}\xspace=45\pm5\,\mbox{${\mathrm{km\,s}}^{-1}$}\xspace$, are quite narrow. In addition \citetads{2002ASPC..279..143C} mention a low surface gravity and strong helium lines, also fitting the picture. The problem is that the system is also an eclipsing binary, which introduces phase-dependent phenomena in the absorption spectrum. Without a more detailed investigation of these variations, it is difficult to judge whether the disk in the system is already that of a classical Be star, i.e., whether it is a slowly outflowing decretion disk as in classical Be stars \citepads{2013A&ARv..21...69R}, or whether it is in the final stages of accretion, or even whether the system needs to be understood in an entirely different framework.

\subsubsection{V658\,Car (HD\,92406)}
V658\,Car was reported as an eclipsing "post-Algol" system with one component being an A0 shell star and the other one "a contracting white-dwarf precursor of $0.28\,M_\odot$" by \citetads{2018BAVSR..67...41H}. The system is reported to have $P=32.1854$\,d, $K_1=10$\,\mbox{${\mathrm{km\,s}}^{-1}$}\xspace, $K_2=80$\,\mbox{${\mathrm{km\,s}}^{-1}$}\xspace, with the higher mass component being cooler and less bright. The pre-subdwarf is estimated to have $\mbox{$T_{\mathrm{eff}}$}\xspace=13$\,kK, which fits well into the parameter space spanned by the other candidates. However, most of the above parameters were derived from photometric data and 19 \mbox{{H}$\mathrm{\alpha}$}\xspace-only spectra from BeSS, that nicely span two orbital cycles. These spectra seem to be similarly complicated as for V505\,Mon, so that a final judgment on the nature of the system should be made only with additional, more modern, spectroscopic extended range and possibly also interferometric observations. Unfortunately, the \mbox{{H}$\mathrm{\alpha}$}\xspace spectra in BeSS do not combine to an RV curve amenable to meaningful quantitative analysis beyond the work by \citetads{2018BAVSR..67...41H}. TESS observations obtained from the archive (see Sect.~\ref{sec:obs:other}) also cover the eclipses and confirm the period.

\subsection{Rejected candidates}
\subsubsection{V1371\,Tau (HD\,36665)}
\begin{figure}[t]
\begin{center}
\includegraphics[angle=0,width=8cm,clip]{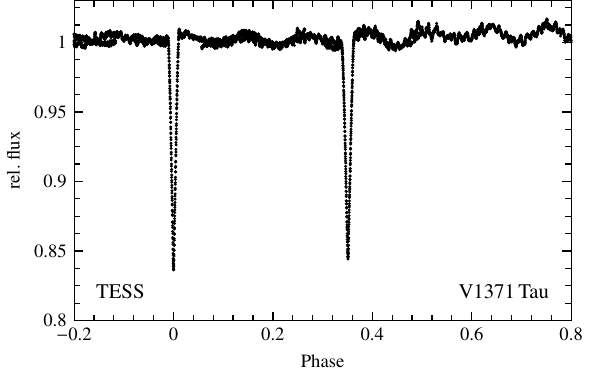}%

\includegraphics[angle=0,width=8cm,clip]{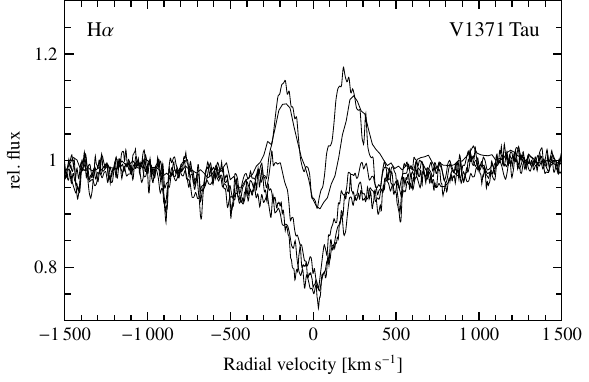}%
\end{center}
\caption[xx]{\label{fig:V1371Tau_TESS}Top: The TESS photometric observations of V1371\,Tau, normalized and folded with the orbital period. Each phase was observed at least twice. Bottom: The \mbox{{H}$\mathrm{\alpha}$}\xspace profiles available from BeSS.}

\end{figure}

\citetads{2020A&A...641A..42B} present V1371\,Tau as a Be star, for which "no indication for a close companion was reported". Inspection of the BeSS spectra, which prior to our campaign consisted of one echelle spectrum and four \mbox{{H}$\mathrm{\alpha}$}\xspace spectra, revealed possible RV variability in \mbox{{H}$\mathrm{\alpha}$}\xspace and \spec{He}{i}{6678}, but the star does not exhibit the signature of narrow absorption lines found in all other candidates. In fact, spectroscopically V1371\,Tau looks like an unremarkable B1\,III star with moderate, but not slow, rotation that sometimes exhibits weak \mbox{{H}$\mathrm{\alpha}$}\xspace emission (Fig.~\ref{fig:V1371Tau_TESS}, bottom). The star was consecutively observed by TESS in sectors 43 to 45. It is a very obvious eclipsing binary, as seen in the top of Fig.~\ref{fig:V1371Tau_TESS}. Over the three months of observation, there are four almost identical, deep and narrow eclipses, signaling a binary of two photometrically similar stars in an eccentric orbit. The period is 33.619\,d with an eccentricity of 0.26.  {This mismatch of showing only one set of spectral lines in a system with (at least) two objects of similar size and luminosity, as is indicated by the eclipses, made it a candidate worth investigating.}

One VLTI/GRAVITY observation was taken, just four days after a conjunction according to the photometric orbit. The system was resolved to be a low-contrast system with two components, at a distance of about 5\,mas. Considering the Gaia DR3 parallax of 0.7\,mas, there is no plausible orbit that would have taken the components to such a separation in just about 100 hours. V1371\,Tau is therefore a hierarchical system of at least three stars, similar to \object{$\nu$\,Gem} \citepads{2021ApJ...916...24K} or HR\,2309 (below), and will be analyzed in greater detail in a different work. 

\subsubsection{HR\,2309 (HD\,44996)}

By visual inspection of spectra in the BeSS database, this star was identified as RV-variable with narrow absorption lines, yet stationary emission.   It was subsequently observed with ARCES and UVES and in a coordinated campaign through BeSS, which confirmed the RV variability of the absorptions and the quasi-stationary Be-type emission. The absorption spectrum is very narrow, similar to LB-1, and not fully resolved by ARCES. Indeed, \citetads{1982ApJS...50...55S} exclaims: "Here is that rare thing: a sharp-lined Be star!" (but inexplicably then lists $\mbox{$v\sin i$}\xspace=50$\,\mbox{${\mathrm{km\,s}}^{-1}$}\xspace in his Table 2). HR\,2309 was, thus, initially considered a high-confidence candidate for the present study.

GRAVITY/VLTI observations show an unambiguous detection of two unresolved sources at 34.4\,mas separation, with a contrast ratio of 44\% to 56\%.  The fainter companion is the Be star, as is evident from the interferometric phase signature across the Br$\gamma$ emission, and is hence dubbed component B. At a parallax of 2.5\,mas, the rather rapid RV changes by up to 50\,\mbox{${\mathrm{km\,s}}^{-1}$}\xspace per day, however, cannot be due to the interferometrically resolved orbit, so that the system must be hierarchical. With newly taken data, the orbit of the narrow lined component could be constrained to a period of about 5.6\,d with $K_{\rm Aa}\sim60\,\mbox{${\mathrm{km\,s}}^{-1}$}\xspace$.  Finally, in \spec{Mg}{ii}{4481}, a third component Ab could be detected spectroscopically, making the system SB3, with two inner narrow-lined components of a mass ratio of about 1:2 and the outer Be-star. Further parameters and analysis of the system will be given in a dedicated future work on Be stars in hierarchical systems.  But the striking spectral similarity to a stripped core should be taken as a cautionary tale, noting that this is quite similar to the architecture originally proposed for HR\,6819 \citepads{2020A&A...637L...3R}.

\subsubsection{{141\,Tau} (HR\,2116, HD\,40724)}

\object{141\,Tau} was discovered as a Be star recently by \citet{2015AJ....149....7C} and is not listed in the BeSS catalog. There are nine ARCES spectra showing a clear composite nature of a B8e spectrum with $\mbox{$v\sin i$}\xspace\approx 120$\,\mbox{${\mathrm{km\,s}}^{-1}$}\xspace and an early A-type spectrum with $\mbox{$v\sin i$}\xspace\approx 3$\,\mbox{${\mathrm{km\,s}}^{-1}$}\xspace. The analysis of these spectra suggest an orbital period of 11.44\,d, with $K=36.5$\,\mbox{${\mathrm{km\,s}}^{-1}$}\xspace and $e=0$ for the narrow lined component, but no detectable motion of the broad-lined component. This would, however, be a very close orbit to form a  stable Be disk within that radius, and on close inspection a third spectral component can just be detected in \spec{Mg}{ii}{4481} at quadrature. This makes the system similar to HR\,2309 as discussed above, and not of the nature this work is searching for. Nevertheless, an interferometric confirmation of a hierarchical nature would be desirable.

\subsubsection{HR\,3195 (HD\,67888, PQ\,Pup)}
In BeSS, there are only one H$\alpha$ and one echelle spectrum available for this object. \citetads{1983A&AS...52..471A} note it as potentially RV variable. Like many other systems studied in that work, \citetads{2019A&A...623A..72K} and \citetads{2021ApJS..254...42B} consider it a star with proper motion anomaly.

HR\,3195 has unresolved narrow absorption lines in the BeSS echelle spectrum, but differs from the other systems by its \mbox{{H}$\mathrm{\alpha}$}\xspace\ emission profile.  The narrow single peak flanked by shoulder inflections \citep[a so-called wine-bottle profile, ][]{1992A&A...262L..17H,1996A&AS..116..309H} suggests that HR\,3195 could be a genuinely single classical Be star viewed pole-on. However, the \mbox{{H}$\mathrm{\alpha}$}\xspace line profile at times shows an additional absorption, making it triple peaked (Fig.~\ref{fig:app:Ha}), which in the authors' experience is usually a good indicator for binarity. On these grounds, HR\,3195 did qualify as a candidate. One interferometric observation did not, however, show a low-contrast companion, but rather no binarity signature was found. As the arguments for a face-on orbit remain valid, this means that either the star is single, or if there is a companion, it must be very faint, and in either case the observed narrow-line absorption spectrum belongs to the Be star.
\section{Discussion\label{sec:discussion}}

\subsection{Statistics of Be stars as binary products}
The BeSS database holds spectra of hundreds of Be stars, many of them observed regularly. Most often these spectra include \mbox{{H}$\mathrm{\alpha}$}\xspace, and sometimes, when echelle instruments were used, also \mbox{{H}$\mathrm{\beta}$}\xspace. However, it is not a homogeneous database in terms of sky and magnitude coverage. For the purpose of statistics, a subsample of BeSS objects has to be defined, that is reasonably complete and from which candidates would have been identified with certainty, namely those stars with observations of sufficient quality and quantity so that the criteria laid out in Sect.~\ref{sec:obs} would have been detected without ambiguity. Since most observers are located in the Northern hemisphere, and have limited telescope power, this sub-sample was chosen as objects listed as classical Be stars in the BeSS database, north of $-30^\circ$, and brighter than or equal to $V=8.0$\,mag. 
Stars north of that declination and brighter than the cutoff typically have more than one hundred spectra in the database, while those south have fewer. It should be kept in mind, however, that for echelle data BeSS counts each order as one spectrum. This gave 328 objects, the vast majority of which has enough observations. Applying the same sub-division as \citetads{2018A&A...609A.108S} (B0 to B2, B3 to B6, and B7 to A1, but see Sect. 4.1 of \citeauthor{2018A&A...609A.108S} for caveats and details), these can be grouped into 136 early-, 96 mid-, and 96 late-type Be stars.

On close inspection, not all of them qualify as classical Be stars (as shown, for instance, in Appendix~\ref{app:compspec}). In turn there are as well newly identified Be stars that are not in the BeSS catalog of classical Be stars, while other candidates, such as V2174\,Cyg or V505\,Mon, are also not in the BeSS database. A detailed discussion of the statistics of Be stars as found in the BeSS database will be published elsewhere, but for now it can be noted that the non-Be stars and the occasional star without sufficient data make between 5 and 10\% of the sample, so that one can estimate about 300 Be stars as the sample population from which candidates are identified. 

Going by the $V$-band magnitudes given in BeSS, this sample then includes three of the confirmed candidates: V447\,Sct, V742\,Cas, and HD\,44637. The latter is probably fainter, but still has a listed magnitude of $V=8$ in the BeSS database, and is thus included. The other confirmed candidates are either too faint (LB-1, V1362\,Cyg) or too southerly (HR\,6819). Of the remaining unconfirmed candidates, HR\,8107 and V505\,Mon would be included in the sample, but the latter is not listed in BeSS. 

{In a typical Be star, almost all light in the visual domain comes from one B-type star. In the binary systems discussed here, this is not true, which will introduce an over-detection bias for a magnitude-limited sample. Assuming the components to be of equal brightness, the found systems will be twice as bright, or detected out to a distance of 1.4 times (considering extinction probably somewhat less) that of the non-binaries. Since Be stars are strongly concentrated to the Galactic plane, i.e., largely two-dimensional, this translates to a correction factor of about two, which is applied in the following discussion. 
}

This means that, in a magnitude-limited sample of Be stars, there are about 0.5--1\% of binary systems in late or immediate post mass-transfer phases. If, instead, also the spectral subtype is taken into account, this incidence more than doubles for early subtypes, and goes to zero for mid and late ones. As a side-note, there is a similar number of hierarchical triples, where the Be star is the outer object, but not the dominant light-source (such as $\nu$\,Gem, \citealt{2021ApJ...916...24K}, or HR\,2309 and 141\,Tau mentioned above). This number can now be compared to other Be-type binaries in the BeSS sample: There are 16 or 17 known or strongly suspected Be+sdOB binaries, which constitute 5\% of the sample
(%
$\varphi$\,Per, 
$\kappa$\,Dra,  
$o$\,Pup,
7\,Vul,
8\,Lac\,A,
28\,Cyg, 
59\,Cyg, 
60\,Cyg, 
HR\,2142, 
HR\,2249, 
HR\,2855, 
HR\,2921, 
HR\,7807, 
QY\,Gem, and 
V1150\,Tau,
see
\citealt{2018ApJ...853..156W,2021AJ....161..248W,2022ApJ...940...86K,2022ApJ...926..213K} and references therein, as well as HR\,1772, Klement et al., in prep., and possibly V1294\,Aql \citealt{2022A&A...666A.136H}).
Further, there are six $\gamma$\,Cas like X-ray stars in the BeSS list, which would be an incidence of 2\% ($\gamma$\,Cas, $\pi$\,Aqr, HR\,2284, HR\,2370, V782\,Cas, and V558\,Lyr, see \citealt{2022MNRAS.510.2286N}). The stars with such $\gamma$\,Cas like X-ray properties form a class that, {among other hypotheses,} has been proposed to consist of Be+WD binaries \citep[see, e.g.,][for the most recent work]{2023ApJ...942L...6G}. {Most of those other hypotheses also consider evolved companions, and a search for radial-velocity variations came up with a considerable number higher number of positive detections and candidates than for other Be stars \citep{2022MNRAS.510.2286N}, so that a binary connection of the $\gamma$\,Cas phenomenon seems likely, even if a single-star magnetic hypothesis cannot be firmly ruled out yet.}

The detection of Be+pre-subdwarf systems in the sample as defined above in this section is likely reasonably complete. The very sharp lines and high radial velocity amplitudes are easy to spot, once these properties have been identified as hallmarks of such systems. The only blind spot are, then, fully face-on orbits, but as HR\,8107, above, suggests even those might be detected. But at a typical value of $K>50$\,\mbox{${\mathrm{km\,s}}^{-1}$}\xspace a system would have to be very close to such an orientation indeed to go unnoticed, {which is probably negligible considering the numbers in the sample}. The $\gamma$\,Cas like X-ray stars are probably also fully complete, as they are easily discovered through X-ray properties at the distances in question. While it is conceivable that there is a currently inactive, i.e., diskless star of this class within the sample, in practice this is probably not the case: all such stars are of early type, which tend to be more active on short timescales, so that an extended inactivity period preventing their detection is unlikely.

However, the list of Be+sdOB systems in the sample is certainly incomplete. They are difficult to detect even with interferometry, and by far not all $\sim300$ stars have been observed with this technique. So, in summary, if one accepts the proposition of $\gamma$\,Cas like X-ray stars as binary products by \citetads{2023ApJ...942L...6G}, this puts the tally of Be stars that are binary products and remain bound to at least 8\% for all Be stars.

It is particularly noteworthy that none of these systems seem to have a truly late-type Be component, they are all early, or at most mid-type Be stars up to about B3, even though the case of 7\,Vul as B4/5+sdO \citepads{2020A&A...639A..32H} and $\kappa$\,Dra as B6IIIe+sdB \citepads{2022ApJ...940...86K} should be kept in mind. While a precise temperature determination is not yet available for most of the Be stars with pre-subdwarf companions, in all the composite spectra the strongest lines from the Be component seem to be of hydrogen and helium, suggesting early subtypes as well. In contrast, not only are late-type Be stars well represented in the BeSS sample, but some of them are also well known to be binaries, such as the B8 star Pleione, for instance \citepads{2010A&A...516A..80N}. Pleione, interestingly, has a highly eccentric orbit of $e=0.7$, whereas all the above, where known, have circular or near-circular orbits: The two most eccentric Be+sdO binaries, 59\,Cyg and 60\,Cyg, have $e=0.14$ and $e=0.2$, respectively \citep{2013ApJ...765....2P,2022ApJ...926..213K}. Considering that, the bound binary products make almost 20\% of early-type Be stars, and nearly none of the mid to late type ones.

Highly eccentric Be binaries, on the other hand, such as $\delta$\,Sco with $e=0.94$, \citepads{2011ApJ...729L...5T}, do exist all over the Be star spectral range.  For binary interaction products, circularization is expected as part of the mass transfer, although this might not be completely justified \citep[see][for a discussion]{2007ApJ...667.1170S,2009ApJ...702.1387S}. But such high eccentricities  indicate that not all early-type, and probably few, if any, late-type Be stars are mass-transfer products, as these would have pre-interaction orbits. This has even been proven for the mid-type Be star binary $\alpha$\,Eri, which has $e=0.73$, where the main sequence companion spectrum is detected and the orbit is too close to fit a stable hierarchical system within that size, that could have undergone mass-transfer \citepads{2022A&A...667A.111K}.

\subsection{Potential progenitor systems}\label{sec:discussion:other}
\citetads{2022MNRAS.516.3602E} report HD\,15124 as a semidetached system of early/mid-B+F type stars with about 5 and 1\,$M_\sun$, respectively, and a period of 5.47\,d.  They model the system as on the path towards an LB-1/HR6819-like configuration, although the total mass of the system seems to be a bit on the low side for that. Such systems are, indeed, not rare and often initially classified as normal Be stars, simply on account of presence and morphology of their Balmer emission. Examples, sorted by increasing orbital period, include 
\object{CX\,Dra} \citep[][B2.5Ve+F5III, $P=6.69603$\,d, $K_{\rm Be}=34\,\mbox{${\mathrm{km\,s}}^{-1}$}\xspace, K_{\rm F}=156\,\mbox{${\mathrm{km\,s}}^{-1}$}\xspace$]{1996A&A...308..799S}
\object{FF\,Cam} \citep[][Be+K, $P=7.788$\,d, $K_{\rm Be}\ll K_{\rm K}=85\,\mbox{${\mathrm{km\,s}}^{-1}$}\xspace$]{2013CEAB...37...67G},
\object{14\,Lac} \citep[][B3e+F9IV, $P=10.0854$\,d, $K_{\rm Be}=25\,\mbox{${\mathrm{km\,s}}^{-1}$}\xspace, K_{\rm F}=158\,\mbox{${\mathrm{km\,s}}^{-1}$}\xspace$]{1997A&A...324..965H,2006A&A...455.1037L}, 
\object{HL\,Lib} \citep[][B9IVe+FIII, $P=24.615$\,d, $K_{\rm Be}\ll K_{\rm F}=84.1\,\mbox{${\mathrm{km\,s}}^{-1}$}\xspace$]{1990PASP..102..312D}, and
\object{HD\,81357}  \citep[][B8e+K?, $P=33.77458$\,d, $K_{\rm Be}\sim10\,\mbox{${\mathrm{km\,s}}^{-1}$}\xspace, K_{\rm K}=82\,\mbox{${\mathrm{km\,s}}^{-1}$}\xspace$]{2019A&A...629A.105K},
as well as the class of double-periodic variables \citep[DPVs, see][]{2016MNRAS.455.1728M}. Not all of those systems are massive enough to form an early type Be star, in particular not the latter two. However, nothing in the statistics excludes that also some late type Be stars are spun up through binary interaction, it is merely the balance of the various spin-up processes that seems to change drastically from early to late type Be stars. 

Which of those will eventually become classical Be stars, however, can only be said with detailed modeling to predict the final orbit and masses of the post-interaction system \citep[see][for an example]{2018A&A...615A..78G}. Also, even though most of the above systems would make the cut considered in the previous subsection as being northern and bright, this is probably not a statistically very meaningful comparison, as the extended donor companion dominates the light from the system, not the B star component. 

In the case of very large rates of mass-loss and gain, such a system might also be identified as of $\beta$\,Lyr type, where the companion is hidden in an obscuring torus, so the basic components of the binary are an evolved, hot donor star and an invisible (but not black hole)
companion, as suggested by \citet{2002ASPC..279..143C} for \object{FY\,Vel} (also see Sect.~\ref{sec:systems:V505Mon}).

\subsection{Accretion flows}

Four systems, LB-1, V742\,Cas, V447\,Sct, and V1362\,Cyg, do show RV-variable H$\alpha$ emission components that trace the RV curve of the donor in phase, but not necessarily with the same amplitude. V447\,Sct shows that also in some \ion{He}{i} lines. 
This is not typical for a Be star disk in a binary. While tidal distortions may lead to density spirals in the disk and some phase locked variability, this is typically observed as a slight variation of the violet-to-red height ratio of the emission peaks, not as a well traceable, strong additional single emission peak going back and forth \citep{2018MNRAS.473.3039P}.
Since such a peak is also not expected in a fully detached binary, this suggests that there might be some limited mass transfer either still ongoing in those systems, or the situation has reversed and the former donor is now re-accreting material from the Be disk.

\subsection{Rotation}
\subsubsection{Pre-subdwarf rotation}
A striking similarity of all non-rejected candidates is the slow rotation of the mass-donor, i.e., pre-subdwarf component. It should be noted, however, that this is not necessarily representative, as the presence of absorption lines with discrepant low width was used as one criterion for the identification of possible candidate systems. \citetads{2022A&A...667A.122S} make the case that the slow rotation of the former donor can only be explained by efficient magnetic angular momentum transport in its stellar interior during the post main sequence expansion. If this is a common property of early-type stars, the ubiquity of slow rotation would be expected. 

The slow rotation can also be misleading, however, as is seen in HR\,2309 above. Together with LB-1, it has the smallest \mbox{$v\sin i$}\xspace, but was shown to likely be a normal main-sequence star in a hierarchical system.

\subsubsection{Be star rotation}
Other than the donor, the mass gainer, i.e., the future Be star, would spin-up during mass transfer.
Be stars are widely agreed to be rapid rotators \citepads{2013A&ARv..21...69R,2016A&A...595A.132Z}. The interior stellar evolution during core-hydrogen burning, as the core contracts, transports angular momentum outwards, resulting in an evolutionary spin-up {in terms of the critical fraction $w=v_{\rm eq.}/v_{\rm crit.}$}. The viscous decretion disk is the most efficient way for the star to shed that angular momentum, and so a disk will form as soon as mechanisms for mass and angular momentum transfer into the disk become available to the star \citepads{2021ApJ...909..149G,
2018MNRAS.476.3555R}. Not all immediate post-interaction systems would have be of the type discussed in this work, as also mergers are a possibility.

This also means that once a B star has entered the Be regime, it is likely to remain a Be star until the end of its main sequence life. This suggests a possible solution for the difference between early and late-type Be stars discussed above: 
The longer main sequence (MS) life time of late-type Be stars may make the single-star evolutionary spin-up route more viable for them, as they may have had enough time to spin up by single-star evolution. This does not mean there are no binary products in late-type Be stars at all, as for instance the cases like \object{KOI-81} or \object{Regulus} show, both rapidly rotating B8 stars with a white dwarf companion, that are, however, not observed as Be stars \citepads{2015ApJ...806..155M,2020ApJ...902...25G}.

The earlier stars, due to their shorter life time on the MS, and possibly as well as due to their higher binarity fraction, would more typically have required the binary evolution path to spin up sufficiently.

\subsection{Surface abundances}

All donors seem to have surface hydrogen, as can be seen when the hydrogen lines are phased with the orbital period. LB-1, V742\,Cas, V2174\,Cyg, and V447\,Sct, show obvious signs of CNO-processing, namely depleted carbon/enhanced nitrogen. V447\,Sct has the most obvious nitrogen enhancement and carbon/oxygen depletion, but no further sign of any enhancement. For HR\,6819, HD\,44637, and V1362\,Cyg the picture is less clear. They do show a CNO-typical pattern in helium, nitrogen, and oxygen, but not the expected strong carbon depletion.
 
\section{Conclusions}

In summary, there are now six systems that have been confirmed as Be binaries with out-of-equilibrium post-overflow companions. Five are spectroscopic binaries in circular orbits (one orbit remaining undetermined until further observations in the coming season), where narrow-lined B-type features dominate the absorption spectrum with high radial velocity, whereas the Be-star emission lines remain largely stationary, i.e. {pointing to either a binary with an extreme mass ratio, and considering the other observational properties, an out-of-equilibrium post-overflow companion}, or a hierarchical system, where the Be star would be an outer component. It is noteworthy that the Be stars in the vast majority of systems that are found as binary products are of early B subtype, suggesting differences in the formation channels of early and late-type Be stars.

Three of those six systems, HR\,6819, V742\,Cyg, and HD\,44637, were positively identified by interferometry as close, low-contrast binaries, without a wider companion. For another three, LB-1, V447\,Sct, and V1362\,Cyg, only the absence of a wider companion was shown interferometrically, but this is sufficient to include them as confirmed cases. 

Three of the candidates, 141\,Tau, HR\,2309, and V1371\,Tau, were identified by interferometry and/or spectroscopy as hierarchical systems and will be analyzed elsewhere. Of the remaining candidates, V2174\,Cyg has all signs of a post-overflow binary, but without interferometric confirmation might still be of a similar kind as HR\,2309. For HR\,8107, HR\,4930, V505\,Mon and V686\,Car further studies are needed, but they remain promising candidates. HR\,3195 could be rejected as being a single star, or at least not having any detectable companion.

This group is sufficient to allow first insights into the statistical properties, first as a subset of Be stars, of which they form about {0.5--1\%} in a magnitude limited sample, but as well as objects in their own right: All systems have zero eccentricity, all show some sign of processed material in their spectra, and all the former donors have rather slow, but except LB-1 not quite zero rotation, while the Be stars have naturally a rapid one.

Further studies will have a closer look at the chemical abundances of the donor stars, and at better constrained orbital parameter, and in particular will make an attempt at obtaining a combined astrometric and radial velocity (SB2) solution for the three interferometrically resolved systems through spectral disentangling.

\begin{acknowledgements}
Part of this work is based on observations collected at the European Organisation for Astronomical Research in the Southern Hemisphere under ESO programmes 
073.D-0274(A), (HR6819 FEROS)
073.C-0337(A), (LS\,MUS FEROS)
2104.D-5024(A), (LB-1 UVES)
0106.D-0994(A), (LB-1 GRAVITY)
0110.D-0400(A), (HD44996 \& HD44673 \& HD36665 UVES)
0110.D-4381(A), (HD44996 GRAVITY)
0110.D-4381(B),  (HD44637 \& HD36665 GRAVITY)
2110.D-5034(A), and 2110.D-5034(B) (HR3195 UVES \& GRAVITY) 
available from the ESO archive at \url{https://archive.eso.org}.
Partly based on observations obtained with the Apache Point Observatory 3.5-meter telescope, which is owned and operated by the Astrophysical Research Consortium.
This work is based upon observations obtained with the Georgia State University Center for High Angular Resolution Astronomy Array at Mount Wilson Observatory.  The CHARA Array is supported by the National Science Foundation under Grant No. AST-1636624 and AST-2034336.  Institutional support has been provided from the GSU College of Arts and Sciences and the GSU Office of the Vice President for Research and Economic Development.
This work has made use of the BeSS database, operated at LESIA, Observatoire de Meudon, France.
BeSS observers: P.~Berardi, E.~Bertrand, C.~Buil, S.~Charbonnel, A.~de~Bruin, V.~Desnoux, J.~Foster,
O.~Garde, J.~Guarro Fl\'o, B.~Heathcote,  F.~Houpert,   R.~Leadbeater,  T.~Lemoult, T.~Lester, M.~Pujol, O.~Thizy.
This research has made use of NASA’s Astrophysics Data System Bibliographic Services.
This research has made use of the SIMBAD database, operated at CDS, Strasbourg, France.
This research has made use of the Jean-Marie Mariotti Center Aspro and SearchCal services. 
This paper includes data collected with the TESS mission, obtained from the MAST data archive at the Space Telescope Science Institute (STScI). Funding for the TESS mission is provided by the NASA Explorer Program. STScI is operated by the Association of Universities for Research in Astronomy, Inc., under NASA contract NAS 5–26555.
This work has made use of data from the European Space Agency (ESA) mission
{\it Gaia} (\url{https://www.cosmos.esa.int/gaia}), processed by the {\it Gaia}
Data Processing and Analysis Consortium (DPAC,
\url{https://www.cosmos.esa.int/web/gaia/dpac/consortium}). Funding for the DPAC
has been provided by national institutions, in particular the institutions
participating in the {\it Gaia} Multilateral Agreement.
SK acknowledges funding for MIRC-X received funding from the European Research Council (ERC) under the European Union's Horizon 2020 research and innovation programme (Starting Grant No. 639889 and Consolidated Grant No. 101003096). JDM acknowledges funding for the development of MIRC-X (NASA-XRP NNX16AD43G, NSF-AST 1909165) and MYSTIC (NSF-ATI 1506540, NSF-AST 1909165).

\end{acknowledgements}

\bibliographystyle{aa}
\bibliography{presubdwarf}

\begin{appendix}

\section{Spectroscopic data}\label{app:spectroscopic}
In this section, the measured radial velocities (RVs) used for the orbital solutions are listed, and some spectra are shown with a wider wavelength coverage than in the main body of the work. For an introduction to the instruments see Sect.~\ref{sec:obs}. RV measurements were made by fitting a single Gaussian absorption profile to the line core. Due to the different resolutions and partly very narrow lines, some stars, like HR\,2309 for \spec{He}{i}{6678}, reveal their blended nature only in the higher-resolution data. In that case, the Gaussian was fitted such that it covered all substructure, as if the line were unresolved. Since measurement errors are difficult to estimate for such a varied set of instruments, and from the data alone, generic values were adopted.  For the combined dynamic solutions, individual weights were reduced by a factor of four for BeSS\,H$\alpha$ and a factor of two for BeSS\,echelle spectra.

\begin{table}[h]
\caption[xx]{\label{tab:V447SctRV}Radial velocity measurements of the \spec{Si}{iii}{4553} line of V447\,Sct.  
}
\begin{center}
\begin{tabular}{lrr}
\hline\hline
Date  & Radial velocity & Source \\
MJD  & [\mbox{${\mathrm{km\,s}}^{-1}$}\xspace] \\
\hline
 58000.1197           &      $-$33.4      &      ARCES \\
 58022.0989           &         35.9      &      ARCES \\
 58028.0666           &         81.3      &      ARCES \\
 58050.0492           &         32.5      &      ARCES \\
 58301.2970           &      $-$65.6      &      ARCES \\
 58375.1115           &         68.1      &      ARCES \\
 58384.1210           &        107.7      &      ARCES \\
 58397.0684           &         70.8      &      ARCES \\
 58673.2039           &        105.7      &      ARCES \\
 58678.2130           &        110.3      &      ARCES \\
 58696.3201           &          2.9      &      ARCES \\
 58707.1337           &      $-$57.1      &      ARCES \\
\hline
\end{tabular}
\end{center}
\end{table}

\begin{table}[h]
\caption[xx]{\label{tab:V1362CygRV}Radial velocity measurements of the \spec{He}{i}{6678} line of V1362\,Cyg.
}
\begin{center}
\begin{tabular}{lrr}
\hline\hline
Date  & Radial velocity & Source \\
MJD  & [\mbox{${\mathrm{km\,s}}^{-1}$}\xspace] \\
\hline
55087.8210           &    84.9      &    BeSS\,echelle    \\
55475.8571           &    58.4      &    BeSS\,H$\alpha$    \\  
56564.8834           &    80.0      &    BeSS\,H$\alpha$    \\  
57219.1265           & $-$85.4      &    BeSS\,H$\alpha$    \\  
57985.8713           &    59.3      &    BeSS\,echelle    \\
58314.9210           &    46.3      &    BeSS\,H$\alpha$    \\  
58381.2072           &    96.6      &    BeSS\,H$\alpha$    \\  
58719.8686           &    88.5      &    BeSS\,H$\alpha$    \\  
58723.8628           &    75.9      &    BeSS\,H$\alpha$    \\  
58725.9755           &    77.3      &    BeSS\,echelle    \\
58732.9037           &    47.2      &    BeSS\,H$\alpha$    \\  
58735.8035           &     1.4      &    BeSS\,H$\alpha$    \\  
58737.8701           & $-$23.8      &    BeSS\,H$\alpha$    \\  
58738.9377           & $-$30.1      &    BeSS\,echelle    \\
58775.8221           &    83.1      &    BeSS\,echelle    \\
58787.7905           &    53.9      &      BeSS\,H$\alpha$    \\  
59069.9488           &    44.0      &    BeSS\,echelle    \\
59092.8965           & $-$95.3      &    BeSS\,echelle    \\
59109.8971           &    60.2      &      BeSS\,H$\alpha$    \\  
59135.8555           & $-$35.0      &    BeSS\,echelle    \\
59140.8355           & $-$67.8      &    BeSS\,echelle    \\
59160.8820           & $-$42.2      &      BeSS\,H$\alpha$    \\  
59161.8537           & $-$17.5      &    BeSS\,echelle    \\
59174.7977           &    86.2      &    BeSS\,echelle    \\ 
59358.4160           &    23.8      &  ARCES    \\    
59364.4184           & $-$39.5      &  ARCES    \\  
59536.1086           & $-$34.6      &  ARCES    \\    
59538.1148           & $-$58.4      &  ARCES    \\    
59540.1212           & $-$71.9      &  ARCES    \\    
59544.0662           & $-$68.3      &  ARCES    \\     
59741.4843           &    71.9      &  ARCES    \\     
59837.2629           & $-$75.5      &  ARCES    \\    
\hline
\end{tabular}
\end{center}
\end{table}

\begin{table}[h]
\caption[xx]{\label{tab:V742CasRV}Radial velocity measurements of the \spec{He}{i}{6678} line of V742\,Cas. 
}
\begin{center}
\begin{tabular}{lrr}
\hline\hline
Date  & Radial velocity & Source \\
MJD  & [\mbox{${\mathrm{km\,s}}^{-1}$}\xspace] \\
\hline
  52139.0222     &   13.9    &  BeSS\,H$\alpha$    \\   
  52236.8472     &$-$80.4    &  BeSS\,H$\alpha$    \\   
  52245.8149     &$-$37.3    &  BeSS\,H$\alpha$    \\   
  52252.8496     &   32.4    &  BeSS\,H$\alpha$    \\   
  52255.7934     &   49.4    &  BeSS\,H$\alpha$    \\   
  52488.1082     &   80.0    &  BeSS\,H$\alpha$    \\   
  52856.1244     &$-$60.7    &  BeSS\,H$\alpha$    \\   
  54041.8086     &   15.9    &  BeSS\,H$\alpha$    \\   
  54371.8752     &$-$38.2    &  BeSS\,H$\alpha$    \\   
  54763.8919     &$-$34.6    &  BeSS\,H$\alpha$    \\   
  55103.9794     &   11.2    &  BeSS\,echelle    \\       
  55458.9127     &   25.6    &  BeSS\,echelle    \\       
  56186.0120     &   18.9    &  BeSS\,H$\alpha$    \\   
  56906.0273     &   64.2    &  BeSS\,echelle    \\       
  57624.0323     &   33.7    &  BeSS\,echelle    \\       
  57752.8851     &   18.4    &  BeSS\,echelle    \\
  58342.9813     &$-$43.1    &  BeSS\,H$\alpha$    \\     
  58796.2345     &   18.0    &  BeSS\,echelle    \\       
  59092.0670     &   41.3    &  BeSS\,echelle    \\       
  59536.0784     &   56.2    &  ARCES    \\    
  59536.2520     &   55.7    &  ARCES    \\    
  59538.0499     &   48.5    &  ARCES    \\    
  59540.2271     &   36.9    &  ARCES    \\    
  59618.1174     &$-$109.6    &  ARCES    \\    
  59624.1090     &$-$80.0    &  ARCES    \\    
  59629.0900     &$-$37.3    &  ARCES    \\  
  59740.4602     &$-$44.0      &  ARCES    \\  
  59824.2356     & $-$4.0      &  ARCES    \\  
  59837.4067     &$-$105.5     &  ARCES    \\  
  59856.3928     & $-$1.3      &  ARCES    \\  
  59865.3794     &   59.3      &  ARCES    \\  
\hline
\end{tabular}
\end{center}
\end{table}

\begin{table}[h]
\caption[xx]{\label{tab:HD44637RV}Radial velocity measurements of the \spec{He}{i}{6678} line of HD\,44637. Early Measurements obtained from BeSS H$\alpha$-region spectra might be unreliable, in particular the high positive value.
}

\begin{center}
\begin{tabular}{lrr}
\hline\hline
Date  & Radial velocity & Source \\
MJD  & [\mbox{${\mathrm{km\,s}}^{-1}$}\xspace] \\
\hline
54907.9002       &     37.75    &  BeSS\,H$\alpha$    \\        
56012.8501       &     84.44    &  BeSS\,H$\alpha$    \\        
56394.8729       &      8.99    &  BeSS\,H$\alpha$    \\        
56701.9626       &  $-$16.17    &  BeSS\,H$\alpha$    \\        
59123.1893       &  $-$11.24    &  BeSS\,echelle    \\        
59940.9157       &  $-$36.39     &  BeSS\,echelle    \\        
59948.9259       &  $-$35.03     &  BeSS\,echelle    \\        
59953.2194       &  $-$34.14     &  ARCES    \\
59957.2038       &  $-$39.09     &  ARCES    \\ 
59957.9674       &  $-$37.29     &  BeSS\,echelle    \\        
59965.9706       &  $-$32.80     &  BeSS\,echelle    \\ 
59966.1661       &  $-$31.44     &  ARCES    \\       
59966.9466       &  $-$31.90     &  BeSS\,echelle    \\        
59971.9631       &  $-$28.76     &  BeSS\,echelle    \\  
59974.1632       &  $-$27.86     &  ARCES    \\       
59976.9469       &  $-$26.96     &  BeSS\,echelle    \\        
59982.7880       &  $-$24.70     &  BeSS\,echelle    \\        
59987.9495       &  $-$24.70     &  BeSS\,echelle    \\        
60006.9798       &   $-$7.64     &  BeSS\,echelle    \\ 
60011.0913       &   $-$4.04     &  UVES    \\          
60013.8808       &   $-$8.98     &  BeSS\,echelle    \\        
60018.8889       &      2.25     &  BeSS\,echelle    \\        
60020.8628       &   $-$0.90     &  BeSS\,echelle    \\        
60022.8785       &      2.69     &  BeSS\,echelle    \\        
60027.8621       &      4.05     &  BeSS\,echelle    \\  
60035.0275       &     17.06     &  UVES    \\      
60036.8701       &     19.76     &  BeSS\,echelle    \\
\hline
\end{tabular}
\end{center}
\end{table}

\begin{figure*}[b]
\begin{center}
\includegraphics[angle=0,width=\textwidth,clip]{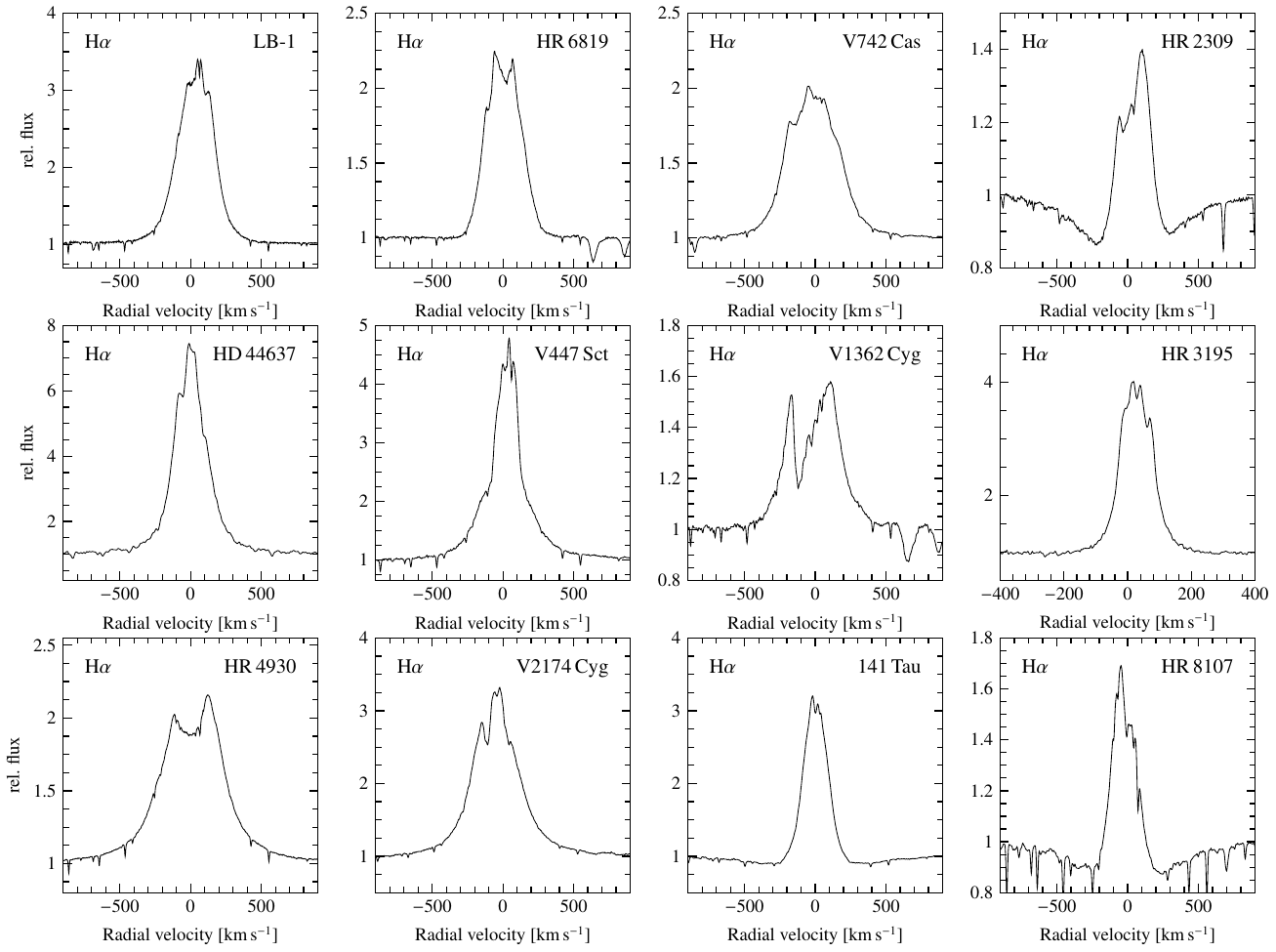}%

\end{center}
\caption[xx]{\label{fig:app:Ha}H$\alpha$ line example spectra of the candidate systems. The velocity scale for HR\,3195 is different.
}

\end{figure*}

\begin{figure*}[t]
\begin{center}
\includegraphics[width=\textwidth]{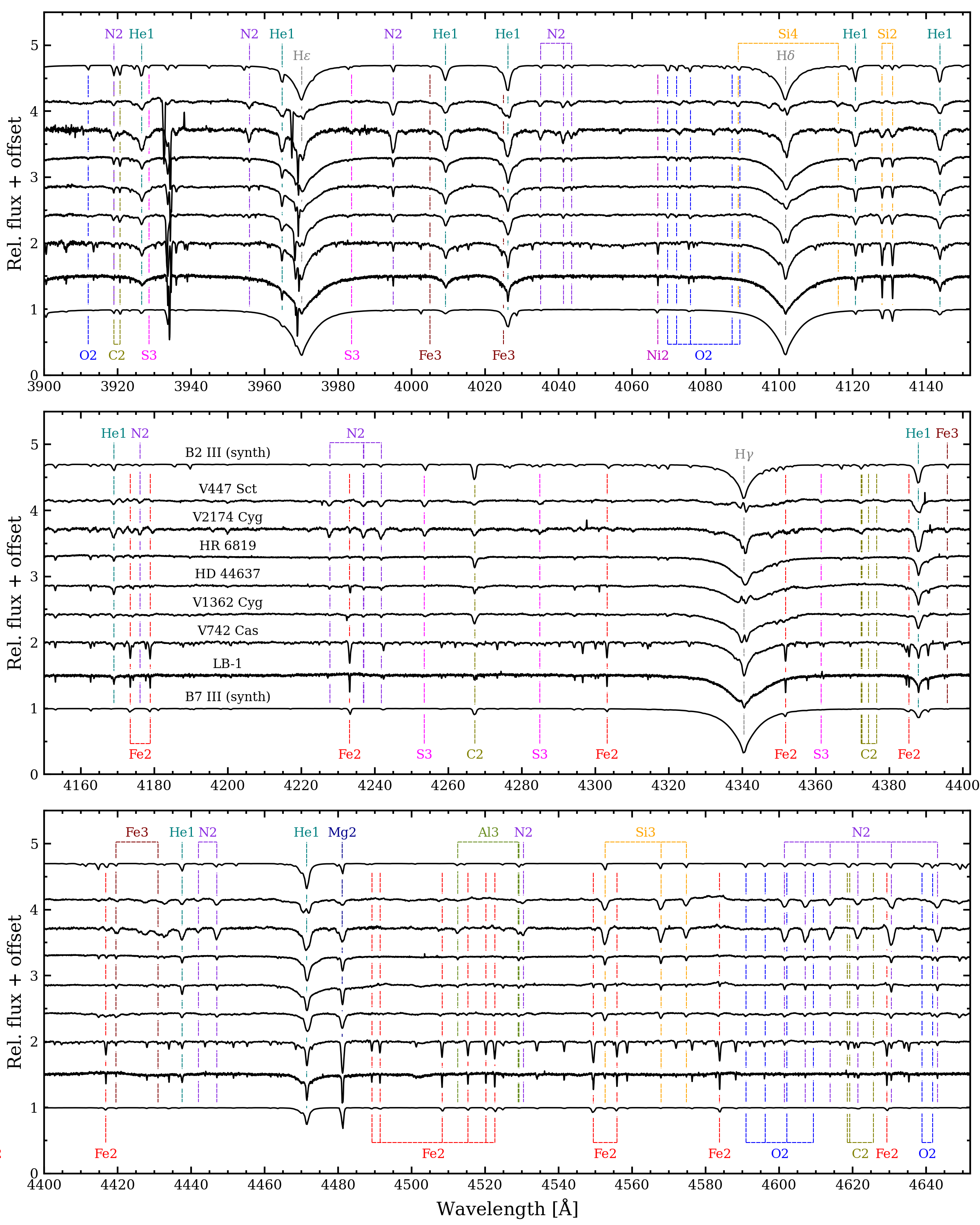}%
\caption[xx]{\label{fig:app:blue}Blue spectral regions for the higher-confidence candidates for which high-resolution spectra are available,  and two synthetic spectra with solar abundances for comparison \citep[computed by][for \mbox{$v\sin i$}\xspace=20\,\mbox{${\mathrm{km\,s}}^{-1}$}\xspace; see there for details]{2018A&A...609A.108S}. Spectra are sorted bottom to top in approximate order of increasing effective temperature, as identified in the middle panel. The RV-variable spectra have been shifted to line up photospheric signatures; as a consequence interstellar and telluric features do not match in wavelength.}
\end{center}
\end{figure*}

\begin{figure*}[t]
\begin{center}
\includegraphics[angle=0,width=18cm,clip]{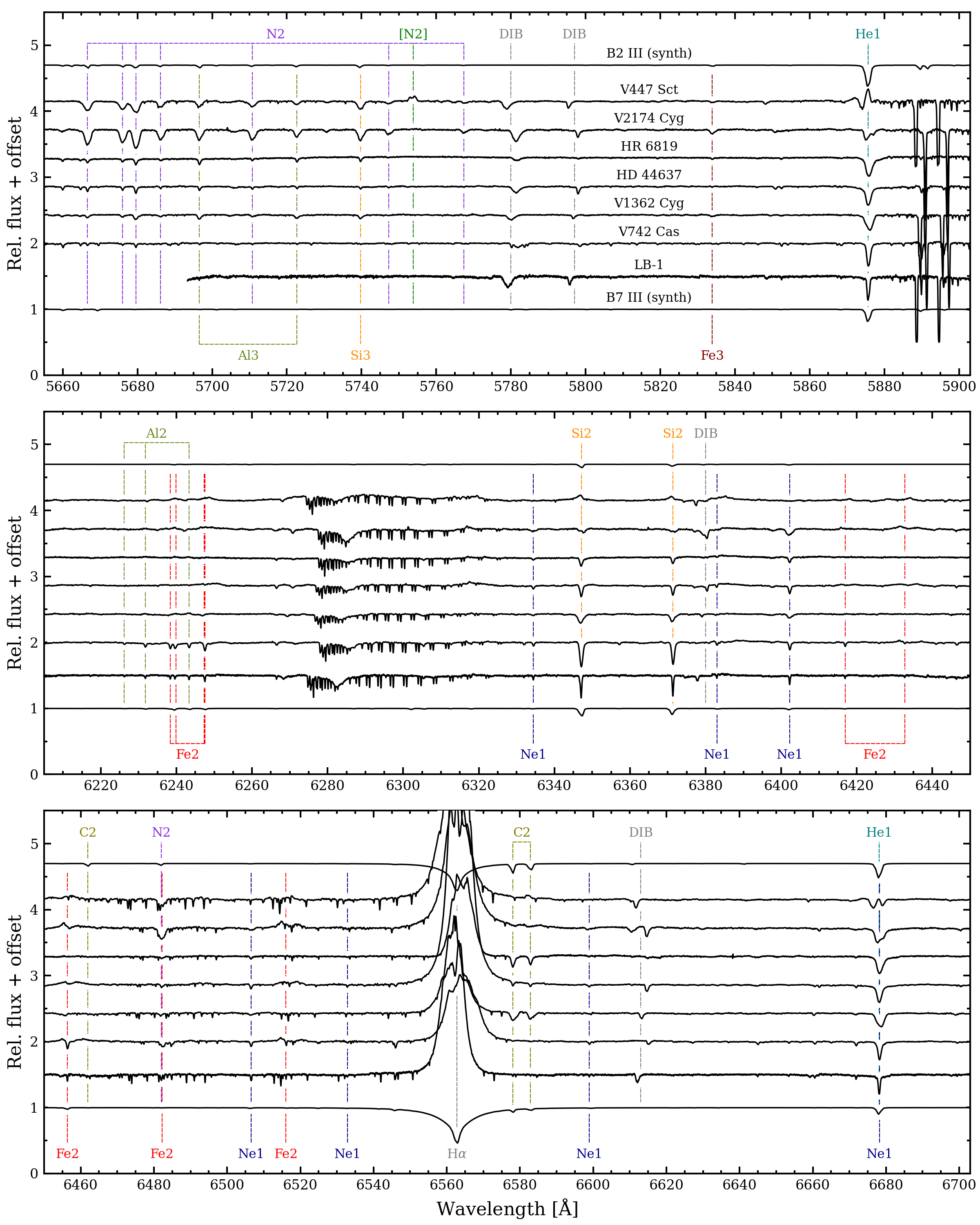}%
\end{center}
\caption[xx]{\label{fig:app:red} Like Fig.~\ref{fig:app:blue}, but for red spectral regions. Strong diffuse insterstellar bands are marked DIB.
}
\end{figure*}

\begin{figure*}[t]
\begin{center}
\includegraphics[angle=0,height=4.7cm,clip]{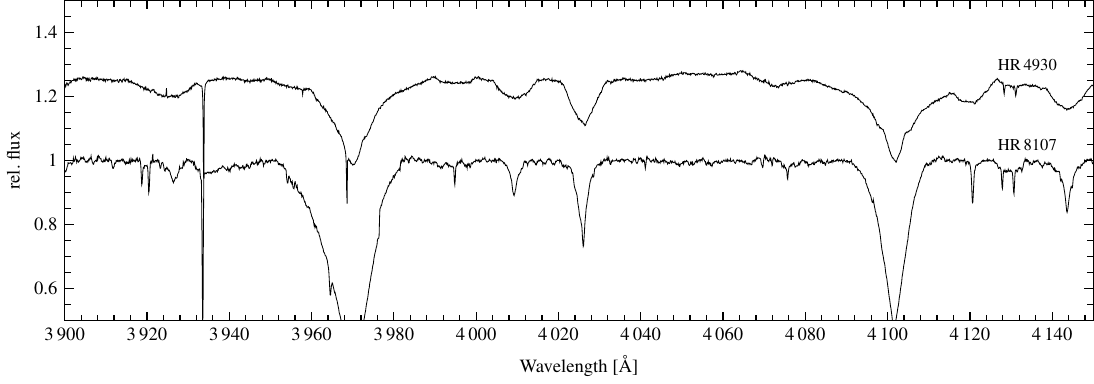}

\hspace{3.25mm}\includegraphics[angle=0,height=4.7cm,clip]{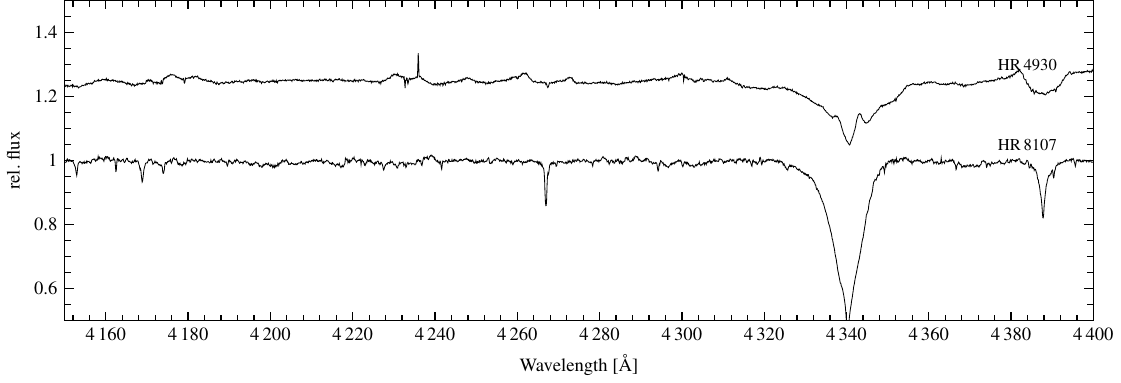}

\includegraphics[angle=0,height=4.7cm,clip]{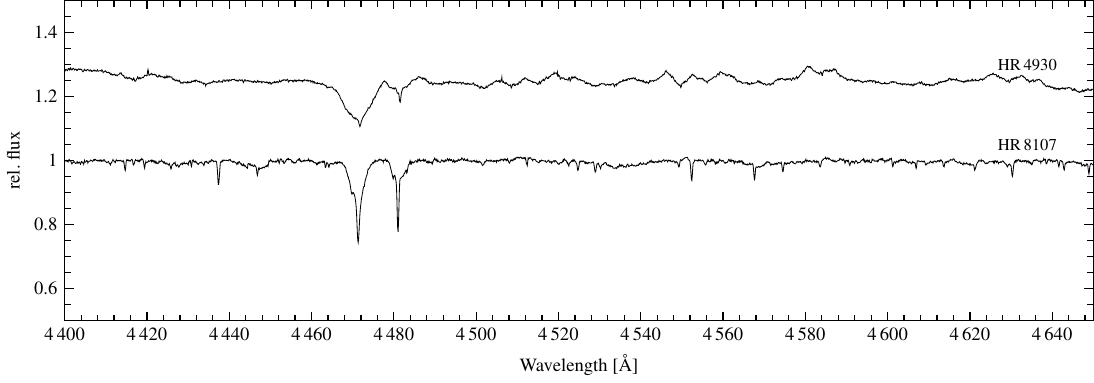}

\includegraphics[angle=0,height=4.7cm,clip]{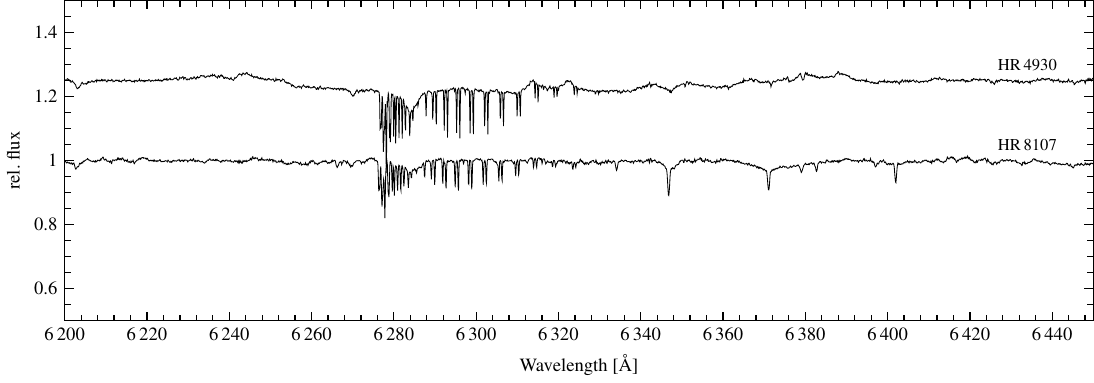}

\hspace{3.5mm}\includegraphics[angle=0,height=4.7cm,clip]{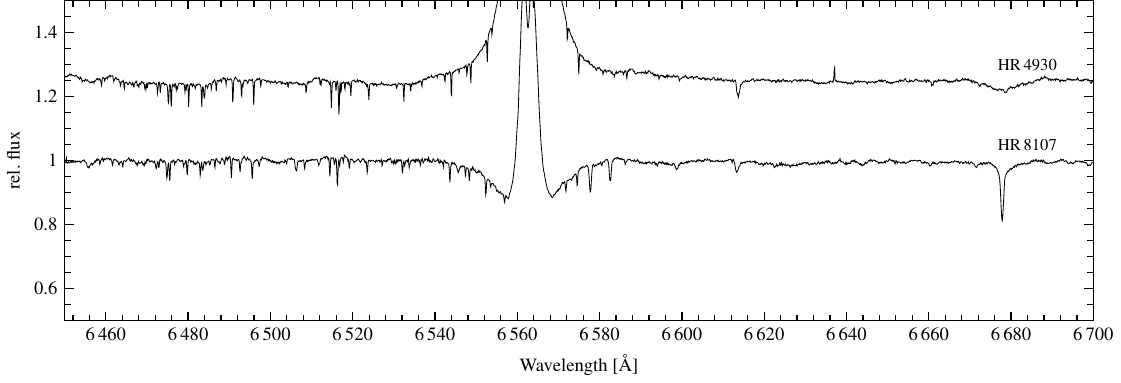}
\end{center}
\caption[xx]{\label{fig:app:cand} Equivalent to Figs.~\ref{fig:app:blue} and \ref{fig:app:red}, but for the unconfirmed candidates HR\,8107 (ARCES) and HR\,4930 (FEROS). No such spectra are available for V505\,Mon and V658\,Car.
}
\end{figure*}

\clearpage
\section{Interferometric data}\label{app:inter}
This section gives tables and example plots for the interferometric observations and their analysis.

\begin{table*}
\caption{Log of interferometric observations}             
\label{tab:interf_log}      
\centering          
\begin{tabular}{l c c c l c} 
\hline\hline       
Target & Instrument & Band/Resolution & MJD$_\mathrm{mean}$ & UT date & Calibrator  \\ 
\hline                    
LB-1 & VLTI/GRAVITY & $K$-low & 59604.090 & 2022 Jan 25 & TYC 1877-265-1 \\
V742~Cas & CHARA/MIRC-X & $H$ & 59732.502 & 2022 Jun 2 & HD 240440\\
V742~Cas & CHARA/MIRC-X & $H$ & 59763.495 & 2022 Jul 3 & HD 240440\\
V742~Cas & CHARA/MIRC-X & $H$ & 59777.408 & 2022 Jul 17 & HD 240440\\
V742~Cas & CHARA/MIRC-X & $H$ & 59818.400 & 2022 Aug 27 & HD 240440\\
V742~Cas & CHARA/MIRC-X & $H$ & 59855.222 & 2022 Oct 3 & HD 240440\\
V742~Cas & CHARA/MIRC-X & $H$ & 59878.245 & 2022 Oct 26 & HD 8992\\
V742~Cas & CHARA/MYSTIC & $K$ & 59732.502 & 2022 Jun 2 & HD 240440\\
V742~Cas & CHARA/MYSTIC & $K$ & 59763.493 & 2022 Jul 3 & HD 240440\\
V742~Cas & CHARA/MYSTIC & $K$ & 59777.408 & 2022 Jul 17 & HD 240440\\
V742~Cas & CHARA/MYSTIC & $K$ & 59818.404 & 2022 Aug 27 & HD 240440\\
V742~Cas & CHARA/MYSTIC & $K$ & 59855.222 & 2022 Oct 3 & HD 240440\\
V742~Cas & CHARA/MYSTIC & $K$ & 59878.245 & 2022 Oct 26 & HD 8992\\
V447~Sct & CHARA/MIRC-X & $H$ & 59763.252 & 2022 Jul 3 & HD 155662\\
V447~Sct & CHARA/MIRC-X & $H$ & 59777.250 & 2022 Jul 17 & HD 166161\\
V447~Sct & CHARA/MYSTIC & $K$ & 59763.251 & 2022 Jul 3 & HD 155662\\
V447~Sct & CHARA/MYSTIC & $K$ & 59777.250 & 2022 Jul 17 & HD 166161\\
V1362~Cyg & CHARA/MIRC-X & $H$ & 59763.438 & 2022 Jul 3 & SAO 68453\\
HR\,2309 & VLTI/GRAVITY & $K$-high & 59928.199 & 2022 Dec 15 & HD~44760 \\
HD\,44637 & VLTI/GRAVITY & $K$-high & 59928.279 & 2022 Dec 15 & HD~44414 \\
V1371\,Tau & VLTI/GRAVITY & $K$-high & 59964.068 & 2023 Jan 20 & HD~245813 \\
HR\,2309 & VLTI/GRAVITY & $K$-high & 59972.055 & 2023 Jan 28 & HD~44760 \\
HR\,3195 & VLTI/GRAVITY & $K$-high & 59989.215 & 2023 Feb 14  & HD~67509 \\
HD\,44637 & VLTI/GRAVITY & $K$-high & 59997.105 & 2023 Feb 22 & HD~44414 \\
\hline                  
\end{tabular}
\end{table*}

\begin{table}[t]
\caption{Interferometric calibrators and their uniform disk (UD) angular diameters}
\label{tab:calibrators}
\centering          
\begin{tabular}{l c c} 
\hline\hline
Calibrator & UD diam. ($H$) & UD diam. ($K$) \\
 & [mas] & [mas]  \\
 \hline                    
TYC 1877-265-1 & $0.037\pm0.001$	& $0.037\pm0.001$ \\
HD 240440 & $0.371\pm0.008$	& $0.373\pm0.008$ \\
HD 8992 & $0.353\pm0.010$	& $0.354\pm0.010$ \\
HD 155662 & $0.431\pm0.012$	& $0.433\pm0.012$ \\
HD 166161 & $0.382\pm0.010$	& $0.384\pm0.010$ \\
SAO 68453 & $0.242\pm0.005$	& $0.243\pm0.005$ \\
HD 44760 & $0.271\pm0.007$	& $0.272\pm0.007$ \\
HD 44414 & $0.185\pm0.005$	& $0.186\pm0.005$ \\
HD 245813 & $0.201\pm0.005$	& $0.202\pm0.005$ \\
HD 67509 & $0.368\pm0.010$	& $0.370\pm0.010$ \\
\hline                
\end{tabular}
\end{table}

\begin{table*}[t]
\caption{Relative astrometric positions and flux ratios of the V742 Cas binary determined from the squared visibilities and closure phases.}             
\label{tab:V742Cas_astrometry}      
\centering          
\begin{tabular}{c c c c c c c c c c} 
\hline\hline       
MJD$_\mathrm{mean}$ & $\rho$ & PA & $\Delta$RA & $\Delta$DEC & $\sigma$-$a$ & $\sigma$-$b$ & $\sigma$-PA & $f$ & Band \\ 
         & [mas]  & [$^\circ$] & [mas] & [mas] & [mas] & [mas] & [$^\circ$] & [\% primary]  & \\
\hline
59732.502 & 0.5974 & 279.544  &$-$0.5891& 0.0991 & 0.0035 & 0.0032 & 200.85 & $49.9^{+0.6}_{-0.7}$ &  $H$ \\
59763.495 & 0.5153 & 87.936   & 0.5150 & 0.0186 & 0.0028 & 0.0027 & 76.79 & $48.3^{+0.3}_{-0.4}$ & $H$ \\
59777.408 & 0.5203 & 320.519  &$-$0.3308& 0.4016 & 0.0029 & 0.0028 & 197.04 & $45.0^{+0.6}_{-0.5}$ & $H$ \\
59818.400 & 0.5619 & 94.429   & 0.5602 &$-$0.0434& 0.0046 & 0.0032 & 115.91 & $40.8^{+0.3}_{-0.4}$ & $H$ \\
59855.222 & 0.3441 & 190.692  &$-$0.0638&$-$0.3381& 0.0046 & 0.0026 & 85.16 & $38.2^{+1.9}_{-2.1}$ & $H$ \\
59878.245 & 0.4160 & 67.428   & 0.3841 & 0.1597 & 0.0054 & 0.0035 & 88.46 & $38.4^{+1.1}_{-1.2}$ & $H$ \\
59732.502 & 0.6164 & 280.909  &$-$0.6053& 0.1167 & 0.0050 & 0.0042 & 144.68 & $55.9^{+0.7}_{-0.8}$& $K$ \\
59763.493 & 0.5159 & 88.496   & 0.5157 & 0.0135 & 0.0035 & 0.0031 & 201.81 & $58.3^{+0.8}_{-0.8}$ & $K$ \\
59777.408 & 0.5187 & 319.372  &$-$0.3377& 0.3937 & 0.0040 & 0.0030 & 152.57 & $54.6^{+1.3}_{-1.4}$ & $K$ \\
59818.403 & 0.5577 & 94.250   & 0.5562 &$-$0.0413& 0.0036 & 0.0031 & 84.47 & $51.9^{+0.7}_{-0.7}$ & $K$ \\
59855.222 & 0.3977 & 195.245  &$-$0.1046&$-$0.3837&  0.0116 & 0.0086 & 87.36 & $32.3^{+3.0}_{-3.4}$ & $K$ \\
59878.245 & 0.4261 & 74.469   & 0.4105 & 0.1141 & 0.0125 & 0.0105 & 74.47 & $37.1^{+3.7}_{-3.9}$ & $K$ \\
\hline                
\end{tabular}
\tablefoot{$\rho$ is the angular separation between the components, PA is the position angle (from North to East), $\Delta$RA and $\Delta$DEC are the companion coordinates relative to the primary, $\sigma$-$a$ and $\sigma$-$b$ are the major and minor axes of the error ellipse, respectively, $\sigma$-PA is the position angle of the error ellipse (from North to East), and $f$ is the secondary-to-primary flux fraction.  The two points taken latest have separations below the nominal resolution, which results in systematically lower $f$ for both instruments. Both components are unresolved and their corresponding diameters were fixed to small values for the binary fit. }
\end{table*}

\begin{figure*}[t]%
\begin{center}
\includegraphics[angle=0,width=12cm,clip]{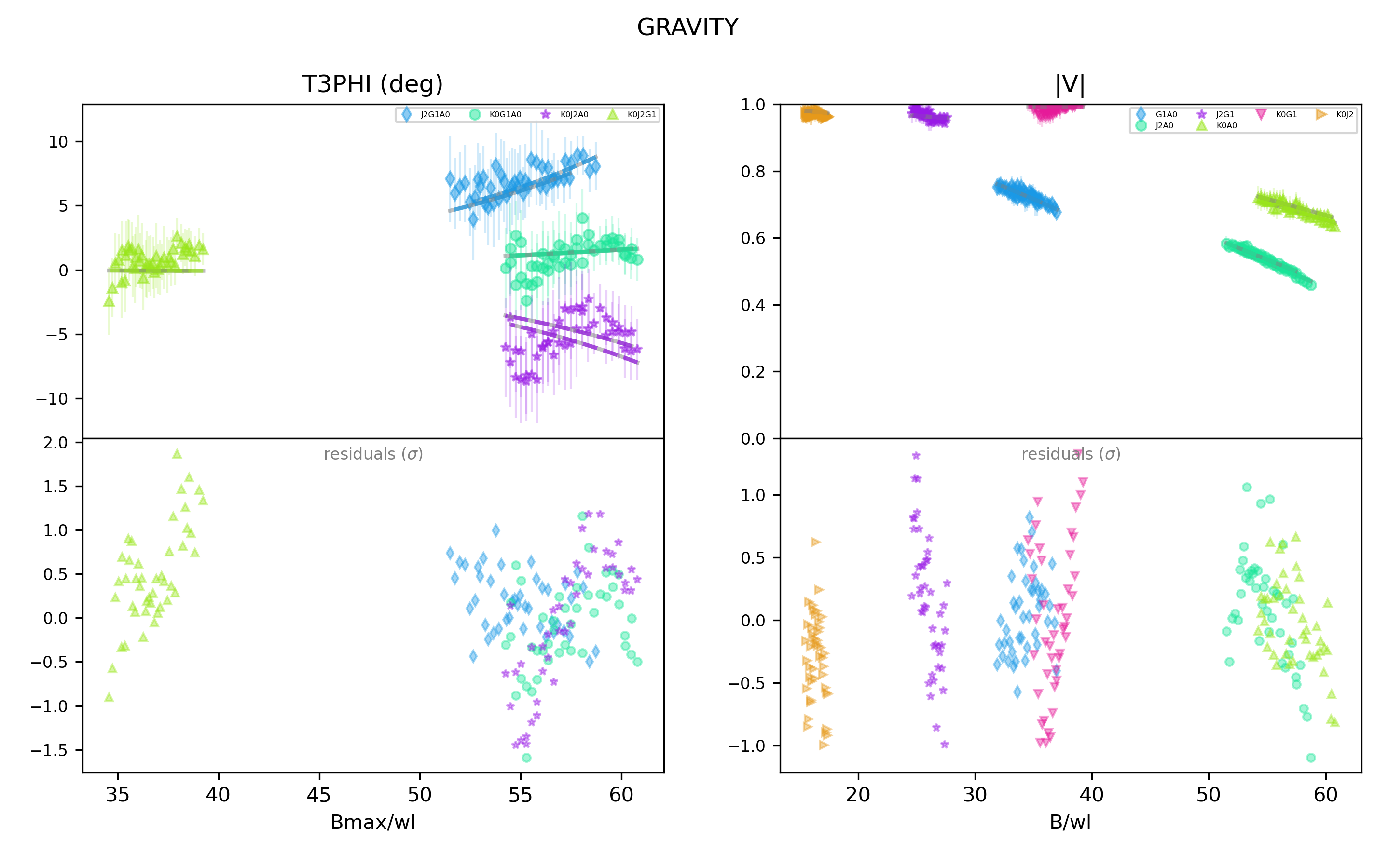}%
\includegraphics[viewport=1 1 285 285,angle=0,width=6cm,clip]{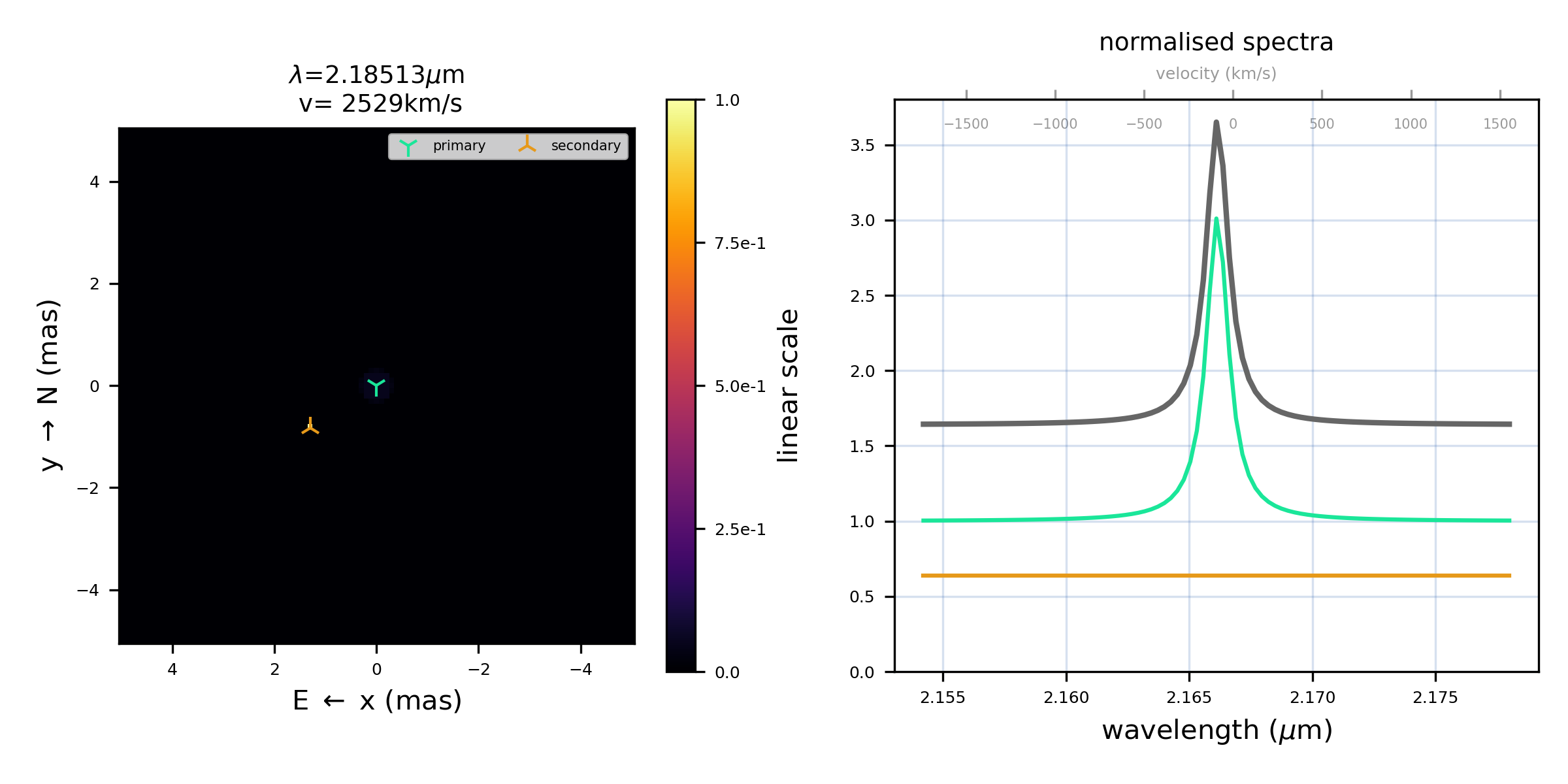}%

\includegraphics[angle=0,width=12cm,clip]{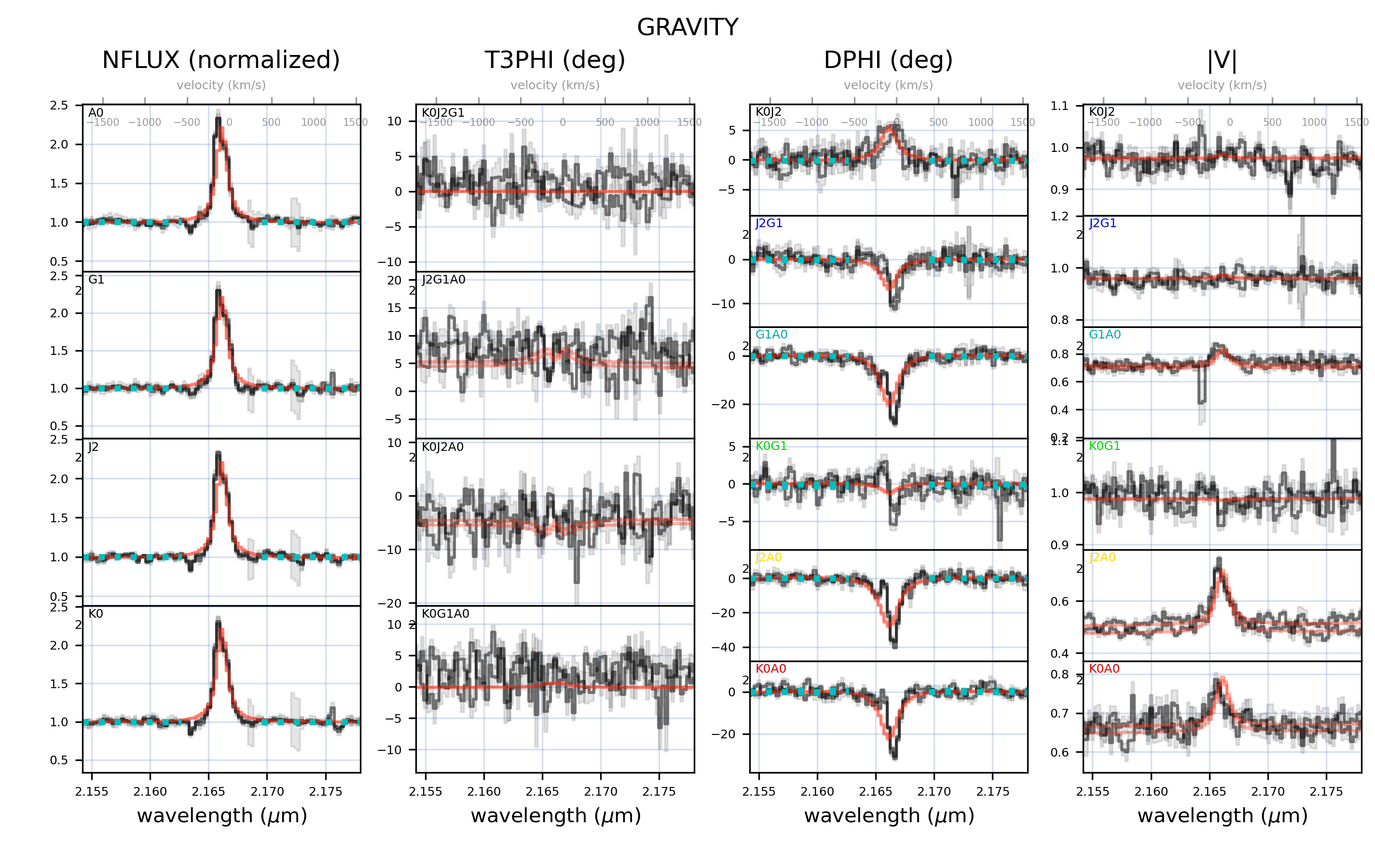}%
\includegraphics[viewport=285 1 570 285,angle=0,width=6cm,clip]{PMOIRED_2022Dec15_SCI_HD44637_SINGLE_SCI_VIS_CALIBRATED.fits_LineModel.png}%
\end{center}
\caption[xx]{\label{fig:app:HD44637_PMOIRED} PMOIRED analysis of the GRAVITY observation of HD\,44637 from 2022 Dec 15. Top left: Binned closure phases (T3PHI) and visibilities (|V|) vs. the spatial frequency, used to determine the relative position and flux ratio of the two binary components. Overplotted is the best-fit binary model (solid lines), with the corresponding residuals  in units of $\sigma$ in the bottom panels. Top right: Binary image on sky with primary being the brighter component.   Bottom left: Full spectral resolution observables including the normalized flux (NFLUX) and differential phase (DPHI) vs. the wavelength across Br$\gamma$ with the best-fit model overplotted in red. Bottom right: Flux contributions from both components across Br$\gamma$. The brighter primary is the Be star showing line emission in Br$\gamma$, while the secondary stripped component shows no discernible line profile in Br$\gamma$.
}

\end{figure*}

\begin{figure*}[t]%
\begin{center}
\includegraphics[angle=0,width=12cm,clip]{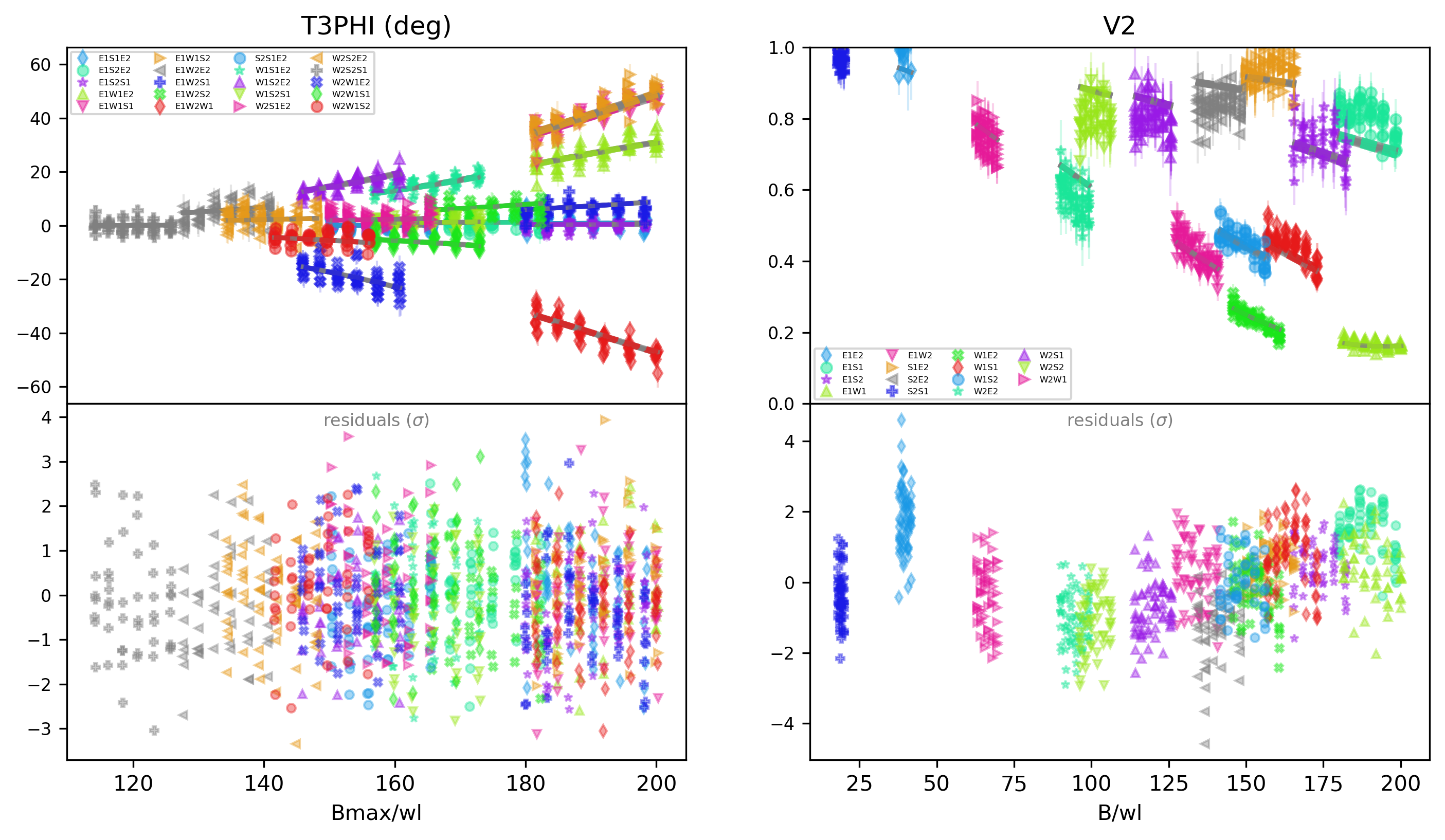}%
\end{center}
\caption[xx]{\label{fig:app:V742Cas_PMOIRED} PMOIRED analysis of MIRC-X observation of V742\,Cas from  2022 Aug 27. Closure phases (T3PHI) and squared visibilities (V2) vs. spatial frequency are overplotted with the best-fit binary model (solid), with the bottom panels showing the corresponding residuals in units of $\sigma$. 
}

\end{figure*}

\section{Other Be stars with composite spectra\label{app:compspec}}

During the inspection of the BeSS database, several objects stood out as being recognized as single, classical Be stars in the database, but showing in fact composite spectra. For some of them, this information could be found in the literature, but not all. Those for which the composite nature seems unpublished are listed below.

\subsection{Be stars with probable late-type companions}
Since the spectra of the following objects are composite, this indicates that the late type spectrum must be of giant-type, as otherwise it would not have been detectable against the B star spectrum, so that they must be low contrast binaries in any case. These might be stars in the process of being stripped, as the ones mentioned in Sect.~\ref{sec:discussion:other}.
\begin{description}
\item[\object{12\,Aur} (HD\,33988)] is a binary, with at least two sets of absorption lines.  In spite of being known as a Be star for nearly a century \citepads{1933ApJ....78...87M}, the literature on this object is rather sparse, which might be due to the proximity of Capella, less than half a degree away.  Whether both spectra are countermoving or the B star spectrum is moving at all is difficult to assess at the data quality.  Certainly one spectrum is rather narrow and much later than B2, of spectral type F, and moves with $K=70$\,\mbox{${\mathrm{km\,s}}^{-1}$}\xspace and a period of 10.7959\,d on a circular orbit.  The other spectrum is near-stationary and that of an early B-type star.  The observed variations of the Balmer emission lines makes it more likely to be an interacting system with ongoing mass transfer, but this is certainly not a final verdict and further observations would be necessary for any firm judgment.

\item[\object{V415\,Aur} (HD\,33461):] The second spectrum is of type F. \spec{He}{i}{6678}, supposedly from the B-type companion, might show signs of RV variability.  There is only one echelle spectrum, so no period analysis can be undertaken, but the more than a dozen \mbox{{H}$\mathrm{\alpha}$}\xspace-spectra show variability typical for an interacting binary.

\item[\object{BD+62 271} (MWC\,14):] A late-type Be-star spectrum, superimposed with a K-type spectrum. The narrow-lined K component is RV variable with several tens of \mbox{${\mathrm{km\,s}}^{-1}$}\xspace amplitude.

\item[\object{HD\,174512}:] There is \mbox{{H}$\mathrm{\alpha}$}\xspace emission, but the spectrum is a clear composite of an early A-type star and a much later narrow-lined component. The BeSS database also includes one FEROS spectrum, where the strong hydrogen lines belonging to the A-type star can be well seen in the Balmer and Paschen sequence.

\item[\object{HD\,55806}] shows a composite spectrum of a mid to late Be type star with Be-typical emission line profiles plus a G-type absorption spectrum. There is one FEROS spectrum in BeSS that confirms the composite nature.
\end{description}

\subsection{Be stars with probable early-type companions}
\begin{description}
\item[\object{HD\,12856}:] The spectrum is reminiscent of HR\,4930 (see Sect.~\ref{sec:systems:HR4930}) and shows broad-lined spectrum of early B-type, superimposed with an unresolved line spectrum of a mid-type B star. There is some indication for low-amplitude RV variability of the narrow lines.  But the number of spectra with sufficient quality is not high enough for a definite conclusion.

\end{description}

\end{appendix}

\end{document}